\documentclass[twocolumn]{aastex6}
\shorttitle{New Low-mass Stars in Orion OB1\MakeLowercase{a}}
\shortauthors{Su\'arez et al. (2017)}

\usepackage{float}
\usepackage{booktabs}
\usepackage{xcolor,colortbl}
\definecolor{Gray}{gray}{0.9}

\begin{document}


\title{New Low-Mass Stars in the 25 Orionis Stellar Group and \\Orion OB1\MakeLowercase{a} Sub-association from SDSS-III/BOSS Spectroscopy}

%% Use \author, \affil, plus the \and command to format author and affiliation 
%% information.  If done correctly the peer review system will be able to
%% automatically put the author and affiliation information from the manuscript
%% and save the corresponding author the trouble of entering it by hand.
%%
%% The \affil should be used to document primary affiliations and the
%% \altaffil should be used for secondary affiliations, titles, or email.

%% Authors with the same affiliation can be grouped in a single
\author{Genaro Su\'arez}
\affil{Instituto de Astronom\'ia, Universidad Nacional Aut\'onoma de M\'exico, Unidad Acad\'emica en Ensenada, Ensenada BC 22860, M\'exico.}

\author{Juan Jos\'e Downes}
\affil{Centro de Investigaciones de Astronom\'ia, Apartado Postal 264, M\'erida, Venezuela \\
Instituto de Astronom\'ia, Universidad Nacional Aut\'onoma de M\'exico, Unidad Acad\'emica en Ensenada, Ensenada BC 22860, M\'exico.}

%% Use the \and command so offset the last author.

\author{Carlos Rom\'an-Z\'u\~niga}
\affil{Instituto de Astronom\'ia, Universidad Nacional Aut\'onoma de M\'exico, Unidad Acad\'emica en Ensenada, Ensenada BC 22860, M\'exico.}

\author{Kevin R. Covey}
\affil{Department of Physics \& Astronomy, Western Washington University, Bellingham WA 98225-9164}

\author{Mauricio Tapia}
\affil{Instituto de Astronom\'ia, Universidad Nacional Aut\'onoma de M\'exico, Unidad Acad\'emica en Ensenada, Ensenada BC 22860, M\'exico.}

\author{Jes\'us Hern\'andez}
\affil{Instituto de Astronom\'ia, Universidad Nacional Aut\'onoma de M\'exico, Unidad Acad\'emica en Ensenada, Ensenada BC 22860, M\'exico.}

\author{Monika G.~Petr-Gotzens}
\affil{European Southern Observatory, Karl-Schwarzschild-Str. 2, 85748 Garching bei M\"unchen, Germany}

\author{Keivan G. Stassun}
\affil{Department of Physics \& Astronomy, Vanderbilt University, Nashville, TN 37235, USA}

\and

\author{C\'esar Brice\~no}
\affil{Cerro Tololo Interamerican Observatory, Casilla 603, La Serena, Chile}


%\altaffiltext{1}{Instituto de Astronom\'ia, Universidad Nacional Aut\'onoma de M\'exico, Ensenada , C.P. 22860, Baja California, M\'exico.}

\begin{abstract}

The Orion OB1a sub-association is a rich low mass star (LMS) region. Previous spectroscopic studies have confirmed 160 LMSs in the 25 Orionis stellar group (25 Ori), which is the most prominent overdensity of Orion OB1a. Nonetheless, the current census of the 25 Ori members is estimated to be less than 50\% complete, leaving a large number of members to be still confirmed. We retrieved 172 low-resolution stellar spectra in Orion OB1a observed as ancillary science in the SDSS-III/BOSS survey, for which we classified their spectral types and determined physical parameters. To determine memberships, we analyzed the H$_\alpha$ emission, LiI$\lambda$6708 absorption, and NaI$\lambda\lambda$8183, 8195 absorption as youth indicators in stars classified as M-type. We report 50 new LMSs spread across the 25 Orionis, ASCC 18, and ASCC 20 stellar groups with spectral types from M0 to M6, corresponding to a mass range of 0.10$\le m/\textrm{M}_\odot \le$0.58. This represents an increase of 50\% in the number of known LMSs in the area and a net increase of 20\% in the number of 25 Ori members in this mass range. Using parallax values from the Gaia DR1 catalog, we estimated the distances to these three stellar groups and found that they are all co-distant, at 338$\pm$66 pc. We analyzed the spectral energy distributions of these LMSs and classified their disks by evolutionary classes. Using H-R diagrams, we found a suggestion that 25 Ori could be slightly older that the other two observed groups in Orion OB1a.

% which represents a $\sim45\%$ of the total member estimated LMSs members.

\end{abstract}




\section{Introduction} \label{sec:intro}

Comprehensive studies of known OB associations in terms of their stellar populations and structural properties provide a firm basis to the understanding of how young star aggregations form and evolve until they eventually disperse to be part of the Galactic disk component. Particularly useful in this respect are the $\sim 10$ Myr {\it fossil star forming regions} \citep[FSFRs; ][]{Blaauw1991}, where one would expect that: \textit{i}) the dust and gas are largely dispersed, and extinction is generally low, permitting the detection of low mass stars (LMSs), \textit{ii}) the members are still spatially concentrated and can be distinguished from the field population, \textit{iii}) only a minority of stars retain optically thick circumstellar disks, \textit{iv}) the active star formation was ceased but its products are still present, \textit{v}) accretion is essentially over, so the objects have attained their final masses, and \textit{vi}) the stars can be considered nearly coeval.

The properties of the LMSs in the FSFRs are of particular importance to understand/clarify the structure and dispersal processes acting on such stellar populations, the circumstellar disk evolution and the possible large-scale dynamical effects. In fact, such studies are not possible solely from the observation of massive stars, as the LMSs do have relatively large pre-main-sequence (PMS) phases, they are characterized by the presence of evolving disks, variable mass accretion and circumstellar and chromospheric activity. Furthermore, they make up the majority of all the stars formed in clusters in terms of number and mass \citep[e.g.,][]{Bastian2010}, and have long lifetimes ($>10^{10}$ yr) in the main sequence (MS).

The Orion OB1 association is one of the largest and nearest star forming regions \citep[e.g. ][]{GenzelStutzki1989,Bally2008,Briceno2008} and contains four distinct sub-associations, which can be distinguished according to their ages and content of gas and dust \citep{Blaauw1964}. With an age of 7-10 Myr and a distance of $\sim$330 pc \citep[e.g.,][]{Briceno2005}, Orion OB1a is the oldest and closest of the Orion OB1 sub-associations. Considering the critical age of 10 Myr, Orion OB1a is an excellent region for studying the early evolution of LMSs. Particularly important is the 25 Orionis stellar group (25 Ori), one of the most numerous and spatially dense 7-10 Myr old populations (r$\sim7$ pc, $\Sigma \sim 128$ stars deg$^{-2}$) known within 500 pc from the Sun \citep{Briceno2007}. 

As mentioned in \citet{Downes2014}, there are other associations of similar age to 25 Ori, but these regions cover relatively extended areas in the sky or are too distant to enable the detection of their LMSs. 25 Ori's unique combination of its distance, age, and area in the sky \citep[360 pc, $\sim7$ Myr, and $\approx 3$ deg$^2$; ][]{Briceno2005,Briceno2007,Downes2014}, makes it a particularly convenient region for studying the population of LMSs. Additionally, 25 Ori is almost free of extinction \citep[$A_V\approx0.30$ mag.; ][]{Kharchenko2005,Briceno2005,Briceno2007,Downes2014}.

Although 25 Ori is a clear spatial overdensity of young LMSs \citep{Briceno2007,Downes2014}, a level of contamination is expected close to 25 Ori from at least two other nearby 10 Myr old stellar groups, identified as ASCC 18 and ASCC 20 by \citet{Kharchenko2005,Kharchenko2013}. Thus, in order to disentangle these groups and identify the complete 25 Ori population, it is necessary to make a study that covers an area into those additional groups, beyond the proposed 25 Ori radius of $1^\circ$ \citep{Briceno2005,Briceno2007}.

Several spectroscopic studies have confirmed, to date, 160 LMS members of 25 Ori \citep{Briceno2005,Briceno2007,Biazzo2011,Downes2014,Downes2015}, which represent about 34\% of its total estimated LMS members \citep{Downes2014}. In this paper, we analyze optical spectra obtained with the SDSS-III/BOSS spectrograph to confirm 50 additional young LMSs in Orion OB1a, of which 22 are inside the 25 Ori's estimated area \citep[$1^\circ$ radius; ][]{Briceno2005,Briceno2007}. This increases the confirmed member sample of 25 Ori by about 20\% in a mass range from 0.1 to 0.6 M$_\odot$. We characterize these new members according to their optical spectral types and spectral features, as well as infrared (IR) photometric signatures of circumstellar disks. The paper is organized as follows: In Section \ref{sec:observations} we describe the optical and IR photometric data, and the optical spectroscopy from the SDSS-III/BOSS survey. In Section \ref{sec:results} we analyze the spectra and describe our results. In Section \ref{sec:singular} we comment on particular objects and in Section \ref{sec:summary} we discuss and summarize the results.

\section{Observations} \label{sec:observations}

\subsection{Optical Photometry}
\label{subsec:Optphot}

The $V$, $R$, and $I$ photometry used in this work was obtained from the CIDA Deep Survey of Orion (CDSO) catalog \citep{Downes2014}, which was constructed by co-adding the multi-epoch optical observations from the CIDA Variability Survey of Orion \citep[CVSO; ][]{Briceno2005}. The sensitivity limits of the CDSO covers the LMS and brown dwarf (BD) population of 25 Ori and its surroundings within the region $79.7^\circ\lesssim\alpha\lesssim82.7^\circ$ and $0.35^\circ\lesssim\delta\lesssim3.35^\circ$. The limiting magnitude of the CDSO photometry in this region is $I_{lim}=22$ and the completeness magnitude is $I_{com}=19.6$ \citep{Downes2014}, enough to assure an $I$ band detection even for the faintest targets of our spectroscopic sample ($I\approx17.0$).

Additionally, we used the $u$, $g$, $r$, $i$, and $z$ photometry from the Sloan Digital Sky Survey (SDSS) catalog \citep[][]{Finkbeiner2004,Ahn2012}. These values are listed in Table \ref{tab:photometry}.

\subsection{IR Photometry}
\label{subsec:IRphot}

The \textit{Z, Y, J, H,} and \textit{Ks} near-IR photometry used in this study was carried out by \citet{PetrGotzens2011} as part of the Visible and Infrared Survey Telescope for Astronomy (VISTA) science verification surveys \citep{Arnaboldi2010}. The 5$\sigma$ limiting magnitudes of the VISTA survey of the Orion star-forming region are $Z=22.5$, $Y=21.2$, $J=20.4$, $H=19.4$, and $Ks=18.6$, which are enough to have VISTA photometry even for the faintest objects in our spectroscopic sample ($J\approx15.0$).

Additionally, we used near-IR photometry from the 2MASS catalog \citep{Skrutskie2006} and mid-IR photometry from IRAC-\textit{Spitzer} \citep[][]{Hernandez2007b} and the WISE All-Sky catalog \citep{Cutri2013}. This IR photometry is listed in Table \ref{tab:photometry}.

\subsection{Spectroscopy}
\label{spectroscopy}

The spectra used in this paper were obtained as part of the Baryon Oscillation Spectroscopic Survey \citep[BOSS; ][]{Dawson2013}, which is one of the four main surveys of SDSS \citep[][]{York2000} in its third phase \citep[SDSS-III; ][]{Eisenstein2011}. The BOSS spectrograph has plates with 1000 fibers of 2'' diameter spanning a field of view of 3.0$^\circ$ in diameter and cover a wavelength range from 3560 \AA\ to 10400 \AA\ with a resolution of R=1560 at 3700 \AA\ and R=2650 at 9000 \AA \citep{Gunn2006,Smee2013}. 

\begin{figure*}[ht!]
	\centering
	\includegraphics[width=1.\textwidth]{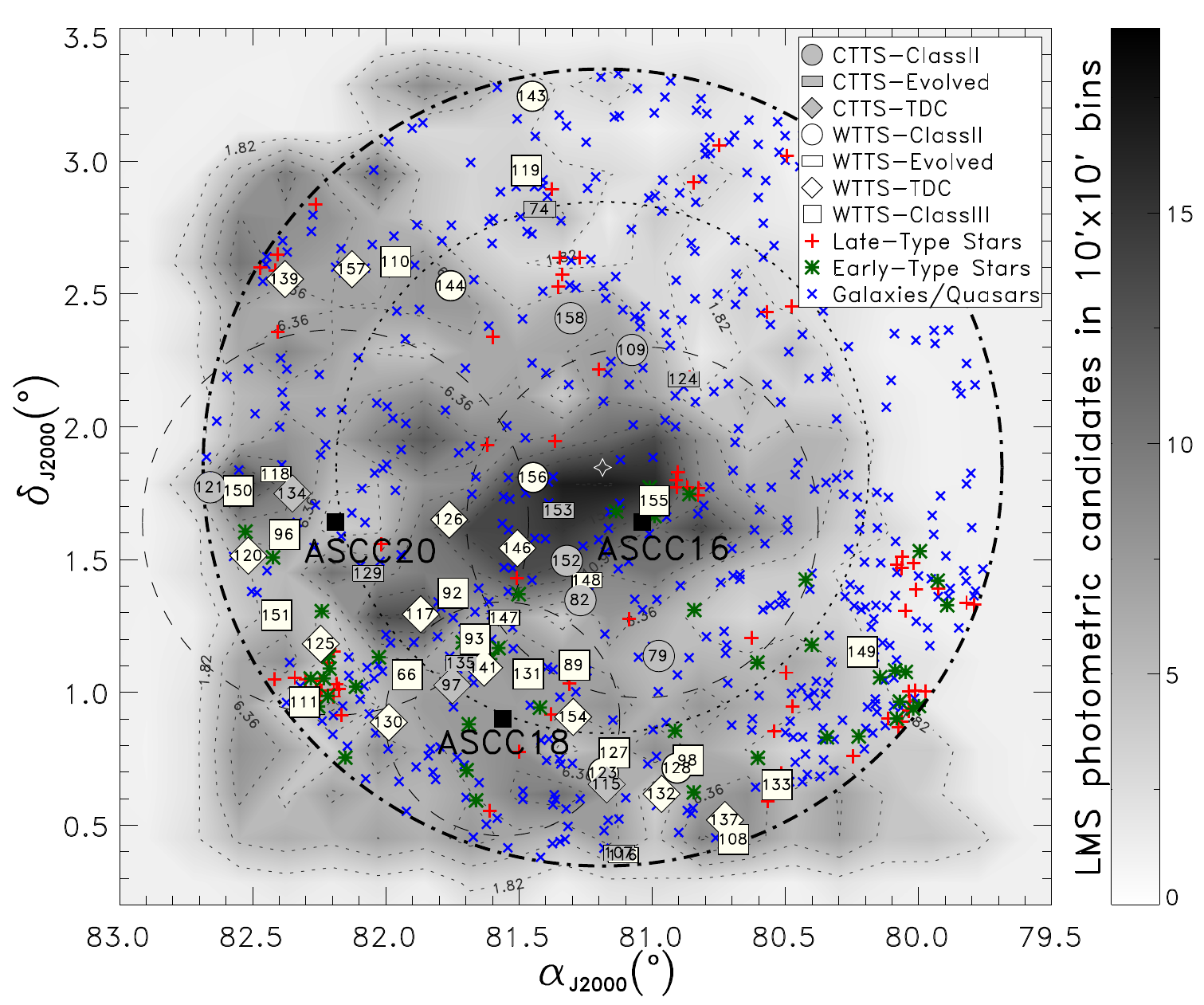}
	\caption{Spatial distribution of the confirmed members of 25 Ori or Orion OB1a classified as CTTSs and WTTSs on the BOSS plate dedicated to 25 Ori (thick dashed-dotted circle); see Sections \ref{subsec:memberships} and \ref{subsec:TTSclass}. The labeled circles, horizontal bars, diamonds, and squares represent YSOs classified as Class II objects, evolved systems, TDC or Class III objects, respectively, as shown in the label box (see Section \ref{subsec:SED}). The dotted circle represents the estimated area of 25 Ori \citep[1$^\circ$ radius; ][]{Briceno2005,Briceno2007}. The red plus signs, green asterisks, and blue cross signs indicate, respectively, stars later than G spectral type, G-type or earlier stars, and the galaxy/quasar samples. The black filled squares and the dashed circles around them represent, respectively, the central position and estimated area of the labeled stellar groups from \citet{Kharchenko2013}. The gray background map and the labeled isocontours indicate the LMS and BD photometric candidate density in 10'x10' bins from \cite{Downes2014}. The white star sign at the center represents the position of the 25 Orionis star.}
	\label{fig:sky}
\end{figure*}

\begin{figure*}[ht!]
	\centering
	\includegraphics[width=1.\textwidth]{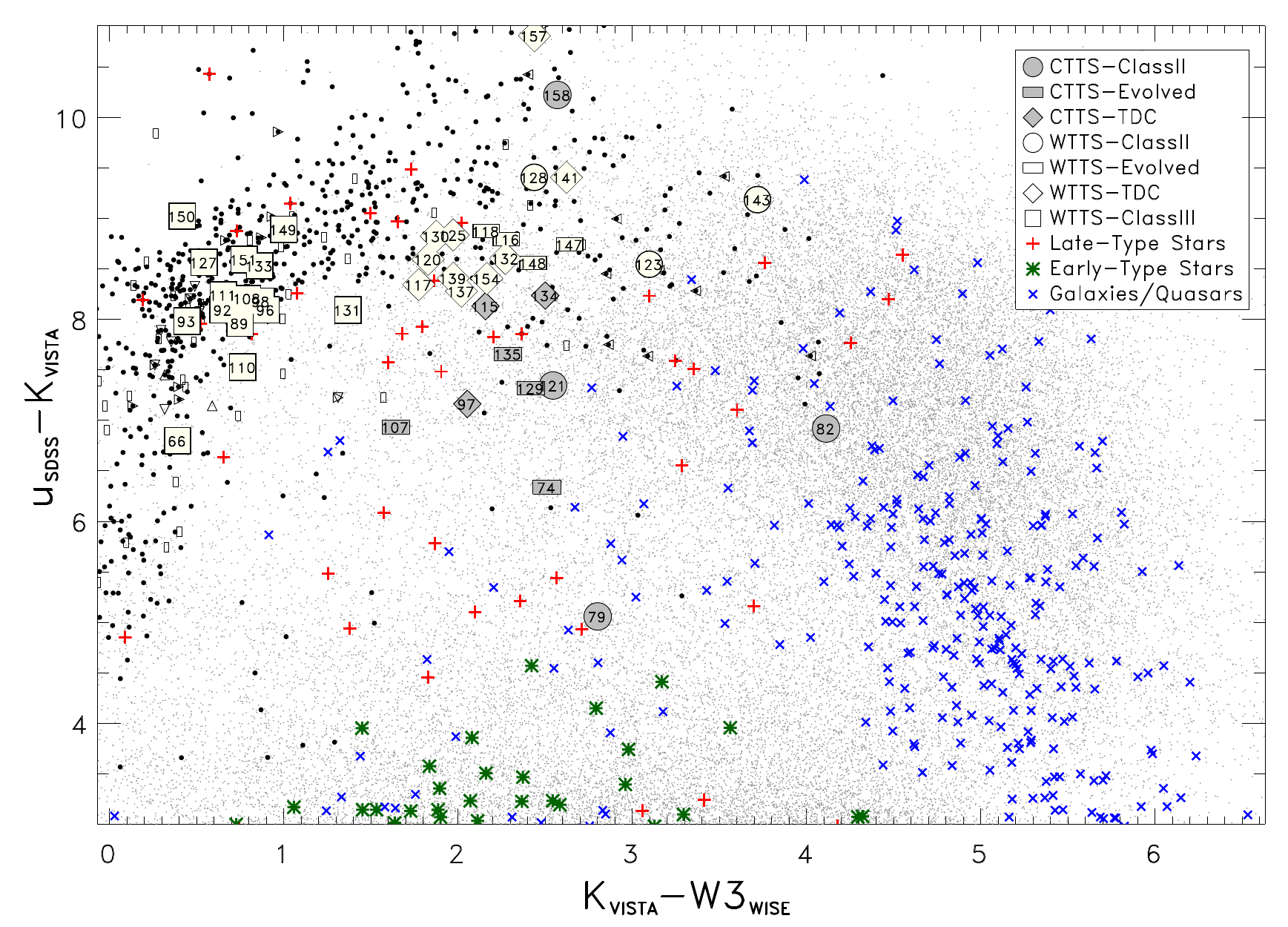}
	\caption{Color-color diagram showing the targets observed in the 25 Ori BOSS plate. The open upward triangles, downward triangles, rightfacing triangles, leftfacing triangles, and vertical bars indicate, respectively, previously confirmed members by \citet{Briceno2005,  Briceno2007, Downes2014, Downes2015}, and Brice\~no \textit {et al.} (in prep.). The black points represent the LMS and BD photometric candidates from \citet{Downes2014}, which were selected with an efficiency of $\sim 86\%$ from color-magnitude diagrams where a bias toward sources with IR excesses is not expected. The gray points show the SDSS+VISTA+WISE detections in the 25 Ori BOSS plate field of view. The rest of the symbols are indicated in the label. Note that the gray labeled symbols represent young stars showing intense H$_\alpha$ emission (see Section \ref{subsec:TTSclass}).}
	\label{fig:CCD_bias}
\end{figure*}

The spectra we analyzed were obtained as part of the {\it Star Formation in the Orion and Taurus Molecular Clouds} ancillary science program \citep{Alam2015}. The plate is centered around the B2 star 25 Orionis ($\alpha_{J2000}= 5^{\rm h} 24^{\rm m} 44^{\rm s}$.8; $\delta_{J2000} = +1^{\circ} 50' 47''.2$) and includes the stellar groups ASCC 16 (25 Ori), ASCC 18, and ASCC 20 \citep{Kharchenko2013}. The object selection process for this plate is described in detail in Section A2 of \citet{Alam2015} and is based on the cataloged optical and IR photometric properties (SDSS, WISE, 2MASS, Spitzer) of the objects. The integration time for the selected objects was 5 x 4500 s, achieving a typical S/N ratio of $\sim$20 for the faintest sources. The observation produced 677 spectra of which only 172 are stellar and the remaining 505 turned out to be spectra of galaxies and quasars (see Section \ref{sec:results}). In Figure \ref{fig:sky} we show the spatial distribution of the targets observed in this plate \citep{Alam2015}, as well as the locations of the different stellar groups from \citet{Kharchenko2013}. In Figure \ref{fig:CCD_bias} we show the $u-K$ vs $K$-W3 color-color diagram including all the targets of the 25 Ori BOSS plate, where we can see that most of these targets have $K$-W3 colors redder than those expected from previously confirmed members. It is important to notice that this target selection implies a bias towards sources with IR excesses (e.g. stars with accretion disks, see Section \ref{subsec:TTSclass}).

The BOSS ancillary science programs made use of the \texttt{v5\_7\_2} version of the \texttt{idlspec2d} pipeline, which, together with \texttt{idlutils}, are the SDSS pipelines used for the data reduction\footnote {\emph{\footnotesize SDSS data processing software is publicly available at \url{http://www.sdss.org/dr12/software/products/}}}. A detailed explanation of the automated classification and the redshift measurements was provided by \citet{Bolton2012}.

The calibrated wavelengths of the BOSS spectra are in the vacuum reference. In order to recognize spectral lines and to use stellar templates to analyze the BOSS spectra, it is needed to convert wavelengths from vacuum to air. We used the IAU standard transformations, as given in \citet{Morton1991}.

\section{Analysis and Results}
\label{sec:results}

\begin{figure*}[ht!]
	\centering
	\includegraphics[width=1.\textwidth]{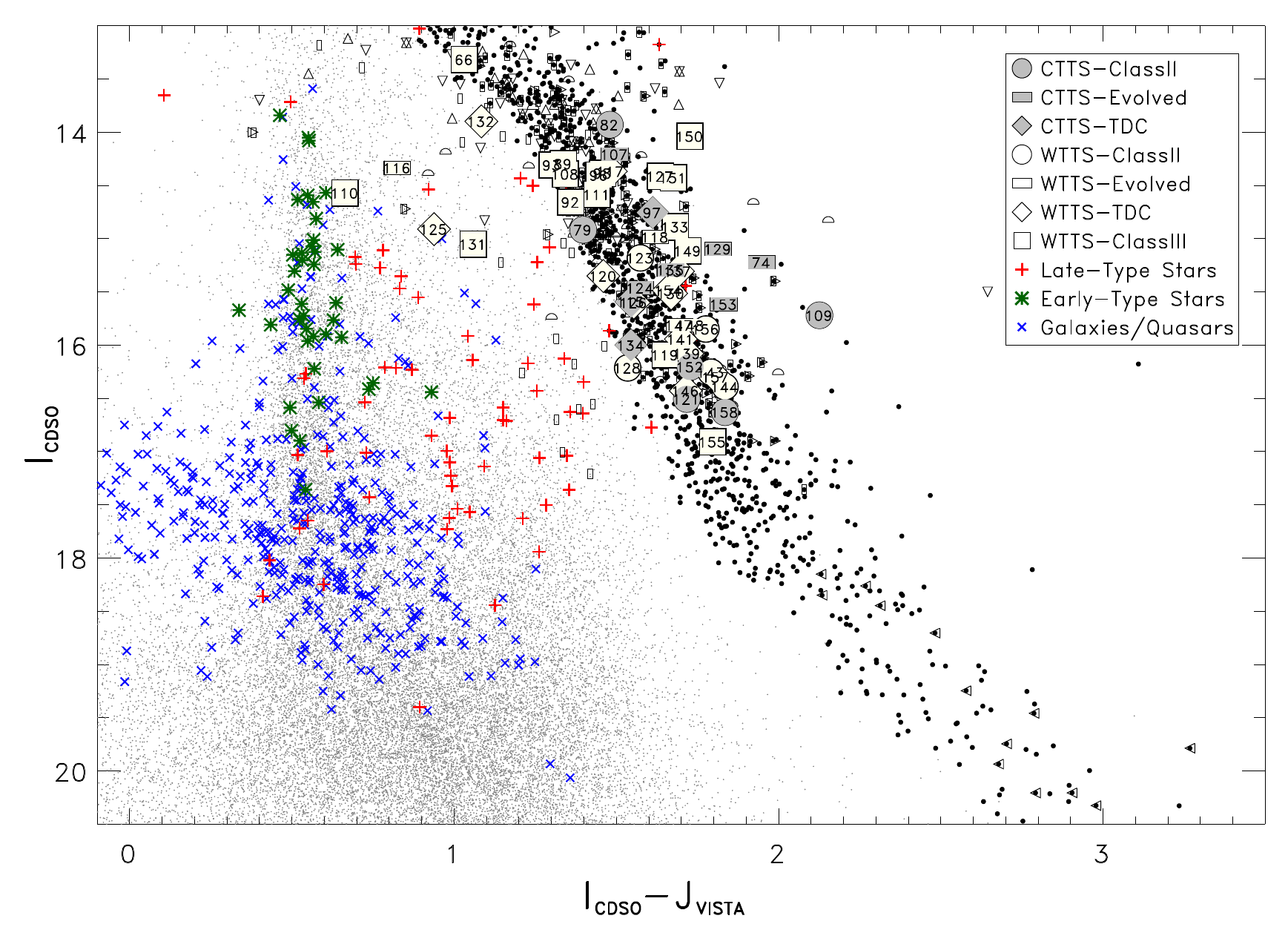}
	\caption{Color-magnitude diagram from the CDSO and VISTA catalogs. The black points and open small symbols are as in Figure \ref{fig:CCD_bias}. The open half circles represent the photometric candidates from \citet{Hernandez2007b}. The gray points show the CDSO+VISTA detections in the 25 Ori BOSS plate field of view. The rest of the symbols are indicated in the label.}
	\label{fig:IvsI-J}
\end{figure*}

In Figure \ref{fig:IvsI-J} we show the $I$ vs $I-J$ color-magnitude diagram for all the targets of the 25 Ori BOSS plate, together with the confirmed members in 25 Ori and Orion OB1a from \citet[][]{Briceno2005, Briceno2007, Downes2014, Downes2015}. We also included the 25 Ori members from Brice\~no \textit{et al.} (in prep.), which were selected according to their position in optical-near-IR color-magnitude diagrams and confirmed on the basis of youth indicators (e.g. H$_\alpha$ emission and NaI$\lambda6708$ absorption) and radial velocities, when available. Additionally, we included the photometric candidates from \citet{Hernandez2007b, Downes2014}. In this diagram, the confirmed members form a very clear locus, nicely separated from the Galactic disk dwarf stars and giant star branches, as well as from extragalactic sources. This shows that the combination of optical and near-IR photometry in color-magnitude diagrams allow for a clear selection of photometric candidates in regions like Orion OB1a \citep[e.g.,][]{Downes2014}.

The BOSS spectra in the DR12 archive are provided with a spectral classification as well as an object classification (star, galaxy or quasar). The stellar templates used for this automated classification are mainly selected from The Indo-US Library \citep{Valdes2004}, which is focused on F- and early G-type stars. The library was supplemented for cool stars by theoretical atmosphere models computed using the MARCS models \citep{Gustafsson2008}. The M-type stellar templates in the database are representative of giant stars with effective temperature down to 3000 K, and a grid resolution of 500 K \citep{Palacios2010}. Thus, these templates are generally not suitable for the spectral type classification of young dwarf stars. Also, the grid resolution in effective temperature of the templates is not enough for the diversity of M-type stars present in our sample. Thus, to determine accurate physical parameters for the BOSS stars (extinction, effective temperature, bolometric luminosity, age, and mass), it is important to verify the SDSS spectral type classification independently. We visually inspected all spectra on the 25 Ori BOSS plate, confirming all objects correctly classified as stars by SDSS, and also identifying those sources for which the SDSS classification was incorrect, either stars classified as galaxies or quasars or, conversely, galaxies or quasars as stars. Through this process, we found 172 stellar spectra out of a total of 677 targets observed on the 25 Ori BOSS plate. The 505 remaining spectra from this plate correspond to either galaxies or quasars (see Figure \ref{fig:sky}).

%\floattable
\begin{splitdeluxetable*}{cccccccccccBcccccccccccccccc}
\tabletypesize{\scriptsize}
\tablewidth{0pt} 
\tablecaption{Photometric catalog of the confirmed members of 25 Ori or Orion OB1a. \label{tab:photometry}}
\tablehead{
ID & $\alpha_{J2000}$ & $\delta_{J2000}$ & \multicolumn{5}{c}{SDSS} & \multicolumn{3}{c}{CDSO} & \multicolumn{5}{c}{VISTA} & \multicolumn{3}{c}{2MASS} & \multicolumn{4}{c}{IRAC} & \multicolumn{4}{c}{WISE} \\
\cmidrule(lr){4-8}
\cmidrule(lr){9-11}
\cmidrule(lr){12-16}
\cmidrule(lr){17-19}
\cmidrule(lr){20-23}
\cmidrule(lr){24-27}
 & & & $u$ & $g$ & $r$ & $i$ & $z$ & $V$    & $R$    & $I$    & $Z$    & $Y$    & $J$    & $H$    & $K$  & $J$ & $H$ & $K$ & CH1 & CH2 & CH3 & CH4 & W1 & W2 & W3 & W4
%		    & ($^\circ$)       & ($^\circ$)       &  & & & & & & & & & & & & & & & & & & & & & & &
} 
\colnumbers
\startdata 
 	66  & 81.919314 & 1.065933 & 18.281    & 15.971    & 14.664    & 14.256    & 13.48     & 15.194 & 14.336 & 13.318 & 13.278     & 12.907     & 12.286     & 11.689     & 11.483     & 12.114     & 11.428     & 11.27      & ...      & ...      & ...      & ...      & 11.242 & 11.183 & 11.092 & 8.305 \\ \arrayrulecolor{Gray}\hline
  	74  & 81.421671 & 2.818307 & 18.326    & 17.989    & 17.294    & 16.394    & 15.679    & 17.935 & 16.772 & 15.223 & 14.415     & 13.89      & 13.275     & 12.563     & 11.989     & 13.585     & 12.862     & 12.353     & ...      & ...      & ...      & ...      & 11.403 & 10.89  & 9.479  & 7.651 \\ \arrayrulecolor{Gray}\hline
  	79  & 80.975436 & 1.136898 & 17.565    & 17.46     & 16.39     & 15.777    & 15.051    & 16.832 & 15.932 & 14.917 & 14.628     & 14.094     & 13.519     & 12.941     & 12.507     & 13.5       & 12.757     & 12.295     & ...      & ...      & ...      & ...      & 11.752 & 11.259 & 9.705  & 7.639 \\ \arrayrulecolor{Gray}\hline
  	82  & 81.269534 & 1.347489 & 18.175    & 17.499    & 16.287    & 15.147    & 14.329    & 16.621 & 15.302 & 13.93  & 13.467     & 12.803     & 12.45      & 11.748     & 11.255     & 12.457     & 11.666     & 11.059     & 10.043   & ...      & 9.22     & ...      & 10.167 & 9.494  & 7.142  & 5.08  \\ \arrayrulecolor{Gray}\hline
  	89  & 81.292164 & 1.102412 & 20.133    & 17.362    & 16.228    & 15.094    & 14.409    & 16.829 & 15.738 & 14.297 & 13.867     & 13.446     & 12.96      & 12.382     & 12.17      & 12.969     & 12.321     & 12.082     & ...      & ...      & ...      & ...      & 12.001 & 11.861 & 11.42  & 8.552 \\ \arrayrulecolor{Gray}\hline
  	92  & 81.747728 & 1.373077 & 20.591    & 18.091    & 16.675    & 15.412    & 14.743    & 17.287 & 16.203 & 14.657 & 14.167     & 13.776     & 13.297     & 12.715     & 12.496     & 13.357     & 12.734     & 12.54      & 12.208   & ...      & 12.209   & ...      & 12.35  & 12.216 & 11.843 & 8.279 \\ \arrayrulecolor{Gray}\hline
  	93  & 81.664785 & 1.200688 & 20.171    & 17.681    & 16.272    & 15.092    & 14.445    & 16.878 & 15.789 & 14.309 & 13.914     & 13.479     & 13.007     & 12.392     & 12.188     & 13.079     & 12.386     & 12.155     & ...      & ...      & ...      & ...      & 12.036 & 11.912 & 11.743 & 8.913 \\ \arrayrulecolor{Gray}\hline
  	96  & 82.379687 & 1.594114 & 20.233    & 17.702    & 16.291    & 15.081    & 14.415    & 16.926 & 15.903 & 14.404 & 13.906     & 13.444     & 12.957     & 12.374     & 12.138     & 13.02      & 12.355     & 12.109     & ...      & ...      & ...      & ...      & 11.991 & 11.852 & 11.239 & 8.369 \\ \arrayrulecolor{Gray}\hline
  	97  & 81.75054  & 1.026903 & 19.279    & 18.115    & 16.778    & 15.467    & 14.674    & 17.569 & 16.325 & 14.755 & 14.133     & 13.661     & 13.143     & 12.501     & 12.116     & 13.18      & 12.47      & 12.102     & ...      & ...      & ...      & ...      & 11.587 & 11.176 & 10.062 & 7.028 \\ \arrayrulecolor{Gray}\hline
  	98  & 80.864815 & 0.743855 & 20.267    & 17.855    & 16.422    & 15.156    & 14.419    & 17.061 & 16.022 & 14.386 & 13.972     & 13.428     & 12.928     & 12.343     & 12.087     & 12.976     & 12.321     & 12.028     & ...      & ...      & ...      & ...      & 11.961 & 11.805 & 11.216 & 8.441 \\ \arrayrulecolor{black}
\enddata
\tablecomments{The complete version of this table is available in the electronic version of this publication.}
\end{splitdeluxetable*}

\subsection{Spectral Classification}

\begin{figure}
	\includegraphics[width=.5\textwidth]{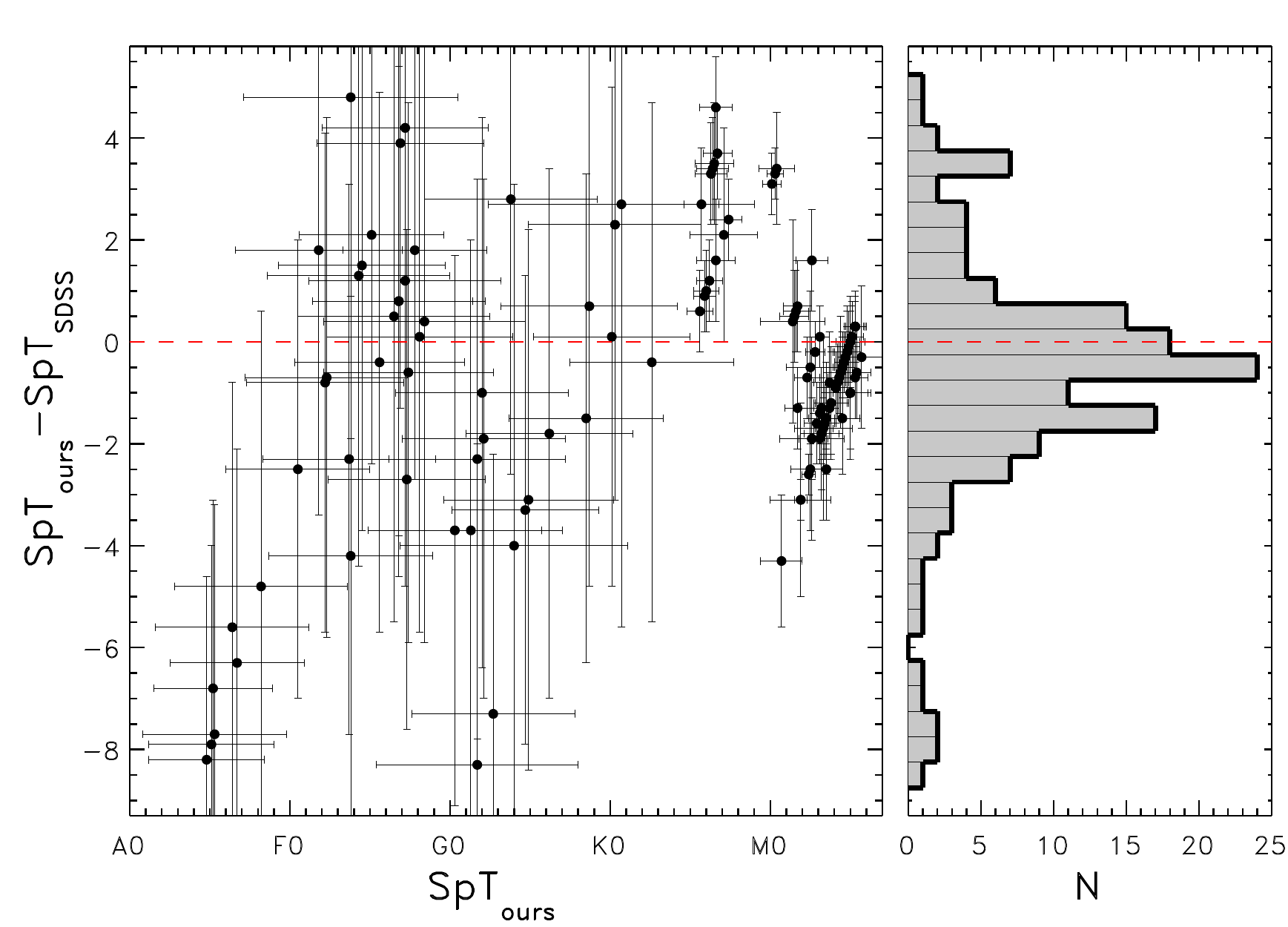}
	\caption{Residuals between our spectral type classification using the SPTCLASS code \citep{Hernandez2004} and the spectral type assigned by the SDSS pipeline. The associated errors are those estimated by the SPTCLASS code because the SDSS classification does not report spectral type uncertainties. The residuals are given in units of spectral sub-classes.}
	\label{fig:sptclass}
\end{figure}

We used the SPTCLASS\footnote {\emph{\footnotesize \url{http://dept.astro.lsa.umich.edu/~hernandj/SPTclass/sptclass.html}}} semi-automatic code (\citealt{Hernandez2004}; extended to classify the M spectral type regime as published in \citealt{Briceno2005}) to derive spectral types for the 172 stars on the 25 Ori BOSS plate. The code uses empirical relations between the equivalent widths of several effective temperature-sensitive spectral features and the spectral types. Particularly, we are interested in the LMS regime, where the SPTCLASS code uses 10 TiO molecular bands in the wavelength range 4775-7150 \AA\ and six VO molecular bands in the mid part of the spectra from 7460 to 8880 \AA. For the LMSs, the typical uncertainties from our spectral type classification are $\pm 0.7$ spectral sub-classes, while these increase up to $\pm 5$ spectral sub-classes for stars earlier than G-type. In Figure \ref{fig:sptclass} we show the residuals between our spectral types using the SPTCLASS code and the spectral types assigned by the SDSS pipeline. Roughly 30\% of the spectra have differences between these two spectral type classifications of more than three spectral sub-classes, especially for the stars with earlier spectral types. There seems to be a trend for stars later than M0, which is due to the fact that most of the M-type stars in our sample are classified as M5 by the SDSS automated classification algorithm. By visually comparing those spectra with the largest spectral type residuals against templates of young and old field stars from \citet{Luhman2000}, \citet{Briceno2002}, \citet{Luhman2003b} \citet{Luhman2004}, and \citet{Kirkpatrick1999}, respectively, we can confirm that our SPTCLASS classification is always more accurate than the SDSS classification. Therefore, for the rest of this work we use the spectral type classification from the SPTCLASS code, which has been extensively used and proven to be accurate and efficient for stars in the spectral type and age ranges considered in this work \citep[e.g.][]{Briceno2007, Hernandez2007b, Downes2014}. This classification covers a spectral type range from A5 to M6, with more than a half of the sample being M-type stars. Spectral types of our confirmed members of 25 Ori or Orion OB1a on the BOSS plate (see Section \ref{subsec:memberships}) are listed in Table \ref{tab:membership}. In Table \ref{tab:field_stars} we list the spectral types of all the stars rejected as members on the 25 Ori BOSS plate.

\subsection{Membership Determination for 25 Ori and \\Orion OB1a}
\label{subsec:memberships}

Once the spectral type was determined for all the stars on the 25 Ori BOSS plate, several diagnoses were applied to determine their memberships, depending on their spectral types.

\subsubsection{M-Type Stars}
\label{subsubsec:M_stars}

The following criteria were used to assign the 25 Ori and Orion OB1a memberships of the M-type stars:

\begin{itemize}
	\item H$_\alpha$ emission.
				
		The detection of strong H$_\alpha$ emission in young LMSs is due to both the chromospheric activity and accretion phenomena that produce narrow symmetric and broader asymmetric H$_\alpha$ line profiles \citep{Muzerolle2005}, respectively. We considered as possible M-type members those stars showing H$_\alpha$ emission. However, because at the age of 25 Ori most of the accreting circumstellar disks have dissipated \citep{Calvet2005}, not only those sources having strong emissions related to accretion are necessary members. Thus, additional criteria are needed to support the memberships of those objects showing weak H$_\alpha$ emission.

	\item LiI$\lambda$6708 absorption.

		For LMSs, the LiI$\lambda$6708 absorption is a well-known indicator of youth \citep{Strom1989, Briceno1998}. It is present in the stellar surface of LMSs because they are fully convective in the PMS phase and the mixing timescale is shorter than the time they need to reach the MS \citep{Soderblom1993}. Therefore, we used the presence of the LiI$\lambda$6708 absorption as an additional membership criterion for the LMSs.

	\item Weak NaI$\lambda\lambda$8183, 8195 doublet in absorption relative to a field star of the same spectral type.

		An additional youth indicator for the M-type stars is the NaI$\lambda\lambda$8183,8195 doublet in absorption, which is sensitive to surface gravity. Since the PMS stars are still contracting, they have lower surface gravity than MS stars with the same spectral type, such that the NaI$\lambda\lambda$8183, 8195 doublet is measurably weaker in PMS sources \citep{Luhman2003a}. We compared the BOSS M-type spectra with templates for young and old field stars with the same spectral type from, respectively, \citet{Luhman2000}, \citet{Briceno2002}, \citet{Luhman2003b} and \citet{Luhman2004}, and \citet{Kirkpatrick1999}, to determine if the NaI$\lambda\lambda$8183,8195 doublet is consistent with the weak absorption expected for a bona fide young star.

\end{itemize}

\begin{figure*}%[H]
	\centering \includegraphics[width=1.\textwidth]{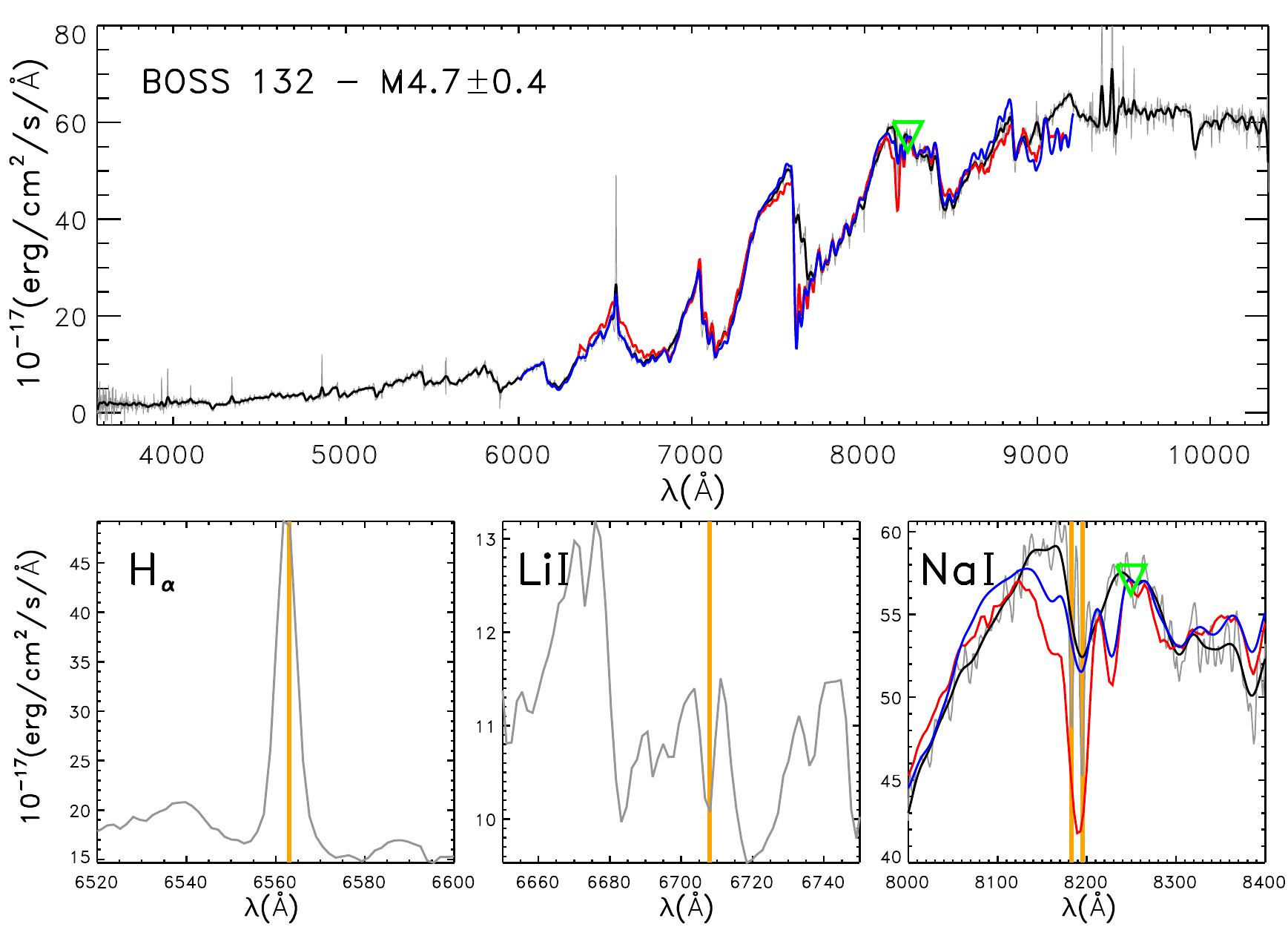}
	\caption{Example of one of the confirmed LMS members in Orion OB1a (member 132 from Table \ref{tab:membership}). \textbf {Upper panel:} BOSS spectrum with the original resolution of 1.4 \AA\ (gray solid line) and with a convolved resolution of 16 \AA\ (black solid line). The stellar templates with the same spectral type as this confirmed member are shown with a resolution of 16 \AA\ for a field star from \citet{Kirkpatrick1999} (red line) and for a young star from the lists of \citet{Luhman2000,Briceno2002,Luhman2003b} and \citet{Luhman2004} (blue line). The green triangle indicates the normalization point of the stellar templates' fluxes, which is located in the pseudo-continuum of the NaI$\lambda\lambda$8183, 8195 doublet of the BOSS spectrum \citep{Schlieder2012}. \textbf {Lower panel:} Enlargements of the H$_\alpha$ emission, LiI$\lambda$6708 absorption and weak NaI$\lambda\lambda$8183, 8195 doublet youth indicators used to assign the memberships of the LMSs. The orange vertical solid lines show the wavelength range of the indicated features. For member 132, the spectrum presents H$_\alpha$ emission, LiI$\lambda$6708 absorption and the NaI$\lambda\lambda$8183, 8195 doublet consistent with the young stellar template. Therefore, this star was confirmed as a young LMS member.}
	\label{fig:membership}
\end{figure*}

Summarizing these criteria, we confirm a M-type star as a 25 Ori or Orion OB1a member if it exhibits any level of H$_\alpha$ emission and either LiI$\lambda$6708 absorption or NaI$\lambda\lambda$8183, 8195 absorption whose profile agrees with a young stellar template of the same spectral type. Based on these criteria, we confirmed a total of 53 members of 25 Ori or Orion OB1a with spectral types from M0 to M6. Three of these members (148, 153, and 156) have already been confirmed as young members by \citet{Downes2014}, so we can report the finding of a total of 50 new confirmed members of 25 Ori or Orion OB1a. In Table \ref{tab:membership} we summarize the membership criteria for these confirmed members. About 87\% of the M-type confirmed members have LiI$\lambda$6708 absorption equivalent widths W(LiI)$>$0.15 \AA\ and the rest have a clear weak NaI$\lambda\lambda$8183, 8195 doublet. The 81\% of the members have the NaI$\lambda\lambda$8183, 8195 doublet in absorption consistent with the young stellar templates and for most of the remaining members the NaI$\lambda\lambda$8183, 8195 is not conclusive but they show a clear LiI$\lambda$6708 absorption. As an example, in Figure \ref{fig:membership} we show the spectrum of the member 157 with enlargements of the H$_\alpha$ emission, LiI$\lambda$6708 absorption and weak NaI$\lambda\lambda$8183, 8195 absorption youth indicators we used to assign its membership.

The 50 new confirmed members represent an increase of $\approx 30\%$ the number of known sub-solar members of 25 Ori or Orion OB1a in the region covered by the considered BOSS plate \citep{Briceno2005,Briceno2007,Hernandez2007b,Downes2014,Downes2015}. Of these members, 22 are inside the \citet{Briceno2005,Briceno2007} estimated area of 25 Ori (1$^\circ$ radius), which represents an increase of $\approx 14\%$ in the number of 25 Ori confirmed members in the sub-solar mass range. Throughout this study, we conservatively worked with the 25 Ori estimated radius of 1$^\circ$, which is greater than the 0.5$^\circ$ radius of the 25 Ori overdensity estimated by \citet{Downes2014}.

\subsubsection{K-Type Stars}
\label{subsubsec:K_stars}

The H$_\alpha$ emission and LiI$\lambda$6708 absorption membership criteria discussed in Section \ref{subsubsec:M_stars} also apply for the K-type stars. Therefore, K-type stars were selected as young stellar objects (YSOs) when presenting H$_\alpha$ emission and LiI$\lambda$6708 absorption. From the 20 K-type stars in the BOSS stellar spectra, we did not confirm any K-type member. None present both LiI$\lambda$6708 absorption and H$_\alpha$ emission (only two stars have weak H$_\alpha$ emission, but those lack LiI$\lambda$6708 absorption).

\subsubsection{Early-Type Stars}
\label{subsubsec:early_stars}

The membership criteria discussed in Section \ref{subsubsec:M_stars} cannot be applied to stars earlier than K-type (43 stars of the sample). In fact, there is not a clear way to confirm these stars as young members with the available information. We checked the position of these stars in the $I$ vs $I-J$ color-magnitude diagram (see Figure \ref{fig:IvsI-J}). We found that not a single early-type star lies within the 25 Ori locus, even when considering the effects of variability. The position of these stars is consistent with the field stars.

The most robust membership diagnostic for stars earlier than K spectral type is their kinematics, though X-ray emission, IR excesses or variability are useful secondary indicators. The spectral resolution of BOSS is not high enough to provide precise radial velocities to determine kinematic memberships for these early type stars, so we checked the other criteria. None of these stars have X-ray counterparts in the 3XMM-DR5 database \citep{Rosen2016} or in the Chandra Source Catalog \citep{Evans2010}. Additionally, none of these stars are high-probability variable stars according to the CVSO catalog \citep{Briceno2005,Mateu2012} or have IR excesses according to the photometric selection of \citet{Cottle2016}, based on 2MASS+WISE photometry and the algorithm developed by \citet{KL2014}. Therefore, the stars earlier than K spectral type on the 25 Ori BOSS plate are likely non-members.

%\newpage
%\floattable
\begin{deluxetable}{ccccc}
\tabletypesize{\tiny}
\tablecaption{Confirmed members and their youth indicators.\label{tab:membership}}
\tablewidth{0pt}
\tablehead{
ID & SpT & WH$_\alpha$ & WLiI$\lambda6708$ & NaI \\
   &     & (\AA)       & (\AA)             &     
\vspace*{-0mm}
}
\decimalcolnumbers
\startdata
	66  & M0.3$\pm$0.5  & -2.062  & 0.4257  & 0   \\ \arrayrulecolor{Gray}\hline
  	74  & M1.5$\pm$1.2  & -161.8  & 0.2514  & -1  \\ \arrayrulecolor{Gray}\hline
  	79  & M1.9$\pm$0.4  & -247.7  & 0.3037  & 1   \\ \arrayrulecolor{Gray}\hline
  	82  & M2.4$\pm$0.4  & -52.95  & 0.2569  & 1   \\ \arrayrulecolor{Gray}\hline
  	89  & M2.8$\pm$0.6  & -5.843  & 0.1607  & 1   \\ \arrayrulecolor{Gray}\hline
  	92  & M3.1$\pm$0.5  & -4.39   & ---     & 1   \\ \arrayrulecolor{Gray}\hline
  	93  & M3.1$\pm$0.5  & -3.541  & 0.1659  & 1   \\ \arrayrulecolor{Gray}\hline
  	96  & M3.2$\pm$0.5  & -4.448  & 0.3134  & 1   \\ \arrayrulecolor{Gray}\hline
  	97  & M3.2$\pm$0.5  & -38.64  & 0.446   & 1   \\ \arrayrulecolor{Gray}\hline
  	98  & M3.2$\pm$0.6  & -5.865  & 0.4586  & 1   \\ \arrayrulecolor{Gray}\hline
  	107 & M3.5$\pm$0.5  & -39.31  & 0.2887  & 1   \\ \arrayrulecolor{Gray}\hline
  	108 & M3.5$\pm$0.6  & -7.876  & ---     & 1   \\ \arrayrulecolor{Gray}\hline
  	109 & M3.5$\pm$1.0  & -41.43  & 0.577   & 0   \\ \arrayrulecolor{Gray}\hline
  	110 & M3.5$\pm$1.1  & -7.721  & 0.2379  & 1   \\ \arrayrulecolor{Gray}\hline
  	111 & M3.7$\pm$0.6  & -4.347  & 0.4181  & 1   \\ \arrayrulecolor{Gray}\hline
  	115 & M4.1$\pm$0.7  & -16.54  & 0.4601  & 1   \\ \arrayrulecolor{Gray}\hline
  	116 & M4.1$\pm$0.7  & -6.188  & 0.4777  & 1   \\ \arrayrulecolor{Gray}\hline
  	117 & M4.2$\pm$0.7  & -6.016  & 0.3093  & 1   \\ \arrayrulecolor{Gray}\hline
  	118 & M4.3$\pm$0.7  & -9.375  & 0.412   & 1   \\ \arrayrulecolor{Gray}\hline
  	119 & M4.4$\pm$1.1  & -14.06  & 0.3332  & -1  \\ \arrayrulecolor{Gray}\hline
  	120 & M4.5$\pm$0.7  & -5.663  & ---     & 1   \\ \arrayrulecolor{Gray}\hline
  	121 & M4.5$\pm$0.7  & -62.18  & 0.0799  & 1   \\ \arrayrulecolor{Gray}\hline
  	123 & M4.5$\pm$0.6  & -6.398  & 0.5469  & 1   \\ \arrayrulecolor{Gray}\hline
  	124 & M4.5$\pm$1.1  & -47.26  & 0.6343  & -1  \\ \arrayrulecolor{Gray}\hline
  	125 & M4.6$\pm$0.6  & -7.454  & ---     & 1   \\ \arrayrulecolor{Gray}\hline
  	126 & M4.6$\pm$0.6  & -8.915  & 0.3241  & 1   \\ \arrayrulecolor{Gray}\hline
  	127 & M4.6$\pm$0.5  & -10.15  & 0.3781  & 0   \\ \arrayrulecolor{Gray}\hline
  	128 & M4.6$\pm$0.5  & -3.17   & 0.5076  & 1   \\ \arrayrulecolor{Gray}\hline
  	129 & M4.7$\pm$0.5  & -74.23  & 0.2364  & 1   \\ \arrayrulecolor{Gray}\hline
  	130 & M4.7$\pm$0.5  & -6.109  & 0.5821  & 1   \\ \arrayrulecolor{Gray}\hline
  	131 & M4.7$\pm$0.5  & -4.569  & 0.5341  & 1   \\ \arrayrulecolor{Gray}\hline
  	132 & M4.7$\pm$0.4  & -12.02  & 0.5568  & 1   \\ \arrayrulecolor{Gray}\hline
  	133 & M4.7$\pm$0.5  & -12.18  & 0.3165  & 1   \\ \arrayrulecolor{Gray}\hline
  	134 & M4.8$\pm$0.4  & -21.34  & 0.1725  & 1   \\ \arrayrulecolor{Gray}\hline
  	135 & M4.8$\pm$0.5  & -59.07  & 0.4103  & 1   \\ \arrayrulecolor{Gray}\hline
  	137 & M4.8$\pm$0.5  & -10.11  & 0.5883  & 1   \\ \arrayrulecolor{Gray}\hline
  	139 & M4.8$\pm$0.9  & -14.6   & ---     & 1   \\ \arrayrulecolor{Gray}\hline
  	141 & M5.0$\pm$0.6  & -9.543  & 0.5009  & 1   \\ \arrayrulecolor{Gray}\hline
  	143 & M5.0$\pm$1.3  & -14.41  & 0.8292  & -1  \\ \arrayrulecolor{Gray}\hline
  	144 & M5.0$\pm$1.1  & -14.36  & 0.8965  & 1   \\ \arrayrulecolor{Gray}\hline
  	146 & M5.1$\pm$0.6  & -15.92  & 0.7128  & 1   \\ \arrayrulecolor{Gray}\hline
  	147 & M5.1$\pm$0.6  & -12.69  & 0.6626  & 1   \\ \arrayrulecolor{Gray}\hline
  	148 & M5.1$\pm$0.7  & -8.659  & 0.6687  & 1   \\ \arrayrulecolor{Gray}\hline
  	149 & M5.1$\pm$0.7  & -11.33  & 0.593   & 1   \\ \arrayrulecolor{Gray}\hline
  	150 & M5.3$\pm$0.6  & -13.43  & 0.354   & 0   \\ \arrayrulecolor{Gray}\hline
  	151 & M5.3$\pm$0.5  & -14.57  & 0.3275  & 1   \\ \arrayrulecolor{Gray}\hline
  	152 & M5.3$\pm$0.7  & -88.18  & 0.5188  & 1   \\ \arrayrulecolor{Gray}\hline
  	153 & M5.3$\pm$0.6  & -32.65  & 0.534   & 1   \\ \arrayrulecolor{Gray}\hline
  	154 & M5.3$\pm$0.6  & -12.34  & 0.7089  & 1   \\ \arrayrulecolor{Gray}\hline
  	155 & M5.3$\pm$0.7  & -14.95  & 1.056   & 0   \\ \arrayrulecolor{Gray}\hline
  	156 & M5.3$\pm$0.8  & -12.88  & ---     & 1   \\ \arrayrulecolor{Gray}\hline
  	157 & M5.4$\pm$0.9  & -7.491  & 0.6515  & 1   \\ \arrayrulecolor{Gray}\hline
  	158 & M5.7$\pm$1.4  & -29.84  & 1.712   & 0   \\ \arrayrulecolor{black}%\hline
\enddata
\tabletypesize{\scriptsize}
\tablecomments{For the NaI$\lambda\lambda$8183, 8195 absorption, the 1, -1, and 0 flags mean, respectively, if the feature is consistent with a young stellar template, an old stellar template or not conclusive.}
\end{deluxetable}

\subsection{Physical Parameters} \label{subsec:parameters}

As described in \citet{Luhman1999}, the $I$ and $J$ bands are preferred to estimate bolometric luminosities of young LMSs because the contamination from UV and IR excess emission is minimal. We used the $I$ band from the CDSO and the $J$ band from VISTA to determine the extinction and bolometric luminosity of our confirmed members.

\subsubsection{Extinctions}
\label{subsubsec:extinction}
The visual extinction ($A_V$) toward each confirmed member was calculated using the observed $I-J$ color, an assumed intrinsic $I-J$ color for a young star of the same measured spectral type and the extinction law from \citet{Fitzpatrick1999}, assuming $R_V=3.1$. The adopted intrinsic $I-J$ color was obtained by interpolating the spectral type in the empirical relations from \citet{KH1995}, \citet{Luhman1999}, \citet{Briceno2002}, and \citet{Luhman2003a}. These relationships were designed to match the \citet{Baraffe1998} tracks, as explained by \citet{Luhman2003b}.

In Table \ref{tab:parameters} we list the extinctions we estimated toward each confirmed LMS member of 25 Ori or Orion OB1a. Removing the two members having the highest unexpected extinctions (member 74 and 109 with $A_V=4.33^{+0.51}_{-0.98}$ and $A_V=3.53^{+0.94}_{-1.01.}$ mag, respectively, see Section \ref{sec:high_extinction}), we obtained a mean extinction $\bar{A}_V$=0.14 mag and a standard deviation $\sigma _{A_V}=0.31$ mag toward the complete sample of confirmed members in 25 Ori and Orion OB1a, which is agreement with the mean extinction $\bar{A}_V$=0.16 mag, obtained by \citet{Downes2015}. If we only consider the confirmed members inside the 25 Ori's estimated area, we obtain $\bar{A}_V$=0.21 mag and $\sigma _{A_V}=0.43$ mag, which is also in agreement with previous extinction estimates in 25 Ori \citep[0.27 mag, 0.28 mag, 0.29 mag, and 0.30 mag by ][respectively]{Kharchenko2005, Briceno2005, Briceno2007, Downes2014}.

\floattable
\begin{deluxetable}{lclclclclclccc}
\tabletypesize{\tiny}
\tablecaption{Physical parameters of the 53 confirmed members.\label{tab:parameters}}
\tablewidth{0pt}
\tablehead{
	ID & $A_\textrm{v}$ & e\_$A_\textrm{v}$ &  T$_{eff}$   & e\_T$_{eff}$  & L           & e\_L          & $m$         & e\_$m$        & age   & e\_age & TTS & Disk Type & Location \\
	   & (mag.)         & (mag.)          & (K)  & (K) & (L$_\odot$) & (L$_\odot$) & (M$_\odot$) & (M$_\odot$) & (Myr) & (Myr)              &     &           &
\vspace*{-2mm}
}
\decimalcolnumbers
\startdata
	66$^a  $    & 0.193 & $^{+0.325}_{-0.203}$ & 3806.5 & $^{+85.5  }_{-72.5 }$ & 0.286 & $^{+0.125 }_{-0.118}$ & 0.573 & $^{+0.038 }_{-0.073}$ & 4.175  & $^{+10.597 }_{-2.387}$  & WTTS & ClassIII & ASCC18/ASCC20 \\ \arrayrulecolor{Gray}\hline
  	74          & 4.33  & $^{+0.513}_{-0.975}$ & 3632.5 & $^{+174.0 }_{-174.0}$ & 0.312 & $^{+0.195 }_{-0.221}$ & 0.425 & $^{+0.101 }_{-0.125}$ & 2.026  & $^{+12.817 }_{-1.338}$  & CTTS & Evolved  & Outside       \\ \arrayrulecolor{Gray}\hline
  	79$^a  $    & 1.355 & $^{+0.183}_{-0.366}$ & 3574.5 & $^{+58.0  }_{-58.0 }$ & 0.111 & $^{+0.028 }_{-0.031}$ & 0.445 & $^{+0.036 }_{-0.045}$ & 8.959  & $^{+5.93   }_{-3.864}$  & CTTS & ClassII  & 25Ori         \\ \arrayrulecolor{Gray}\hline
  	82$^a  $    & 1.299 & $^{+0.426}_{-0.426}$ & 3502.0 & $^{+58.0  }_{-58.0 }$ & 0.276 & $^{+0.09  }_{-0.09 }$ & 0.353 & $^{+0.016 }_{-0.052}$ & 1.79   & $^{+1.686  }_{-0.853}$  & CTTS & ClassII  & 25Ori         \\ \arrayrulecolor{Gray}\hline
  	89$^a  $    & 0.147 & $^{+0.64 }_{-0.538}$ & 3444.0 & $^{+87.0  }_{-87.0 }$ & 0.126 & $^{+0.087 }_{-0.08 }$ & 0.331 & $^{+0.047 }_{-0.031}$ & 4.189  & $^{+10.642 }_{-2.699}$  & WTTS & ClassIII & 25Ori/ASCC18  \\ \arrayrulecolor{Gray}\hline
  	92          & 0.0   & $^{+0.508}_{-0.0  }$ & 3400.5 & $^{+72.5  }_{-72.5 }$ & 0.081 & $^{+0.036 }_{-0.029}$ & 0.306 & $^{+0.053 }_{-0.024}$ & 6.79   & $^{+8.043  }_{-3.529}$  & WTTS & ClassIII & ASCC20        \\ \arrayrulecolor{Gray}\hline
  	93$^a  $    & 0.0   & $^{+0.508}_{-0.0  }$ & 3400.5 & $^{+72.5  }_{-72.5 }$ & 0.117 & $^{+0.074 }_{-0.061}$ & 0.3   & $^{+0.047 }_{-0.019}$ & 3.808  & $^{+11.07  }_{-2.231}$  & WTTS & ClassIII & ASCC18/ASCC20 \\ \arrayrulecolor{Gray}\hline
  	96$^a  $    & 0.33  & $^{+0.482}_{-0.406}$ & 3386.0 & $^{+72.5  }_{-72.5 }$ & 0.119 & $^{+0.052 }_{-0.049}$ & 0.299 & $^{+0.035 }_{-0.034}$ & 3.704  & $^{+7.801  }_{-1.933}$  & WTTS & ClassIII & ASCC20        \\ \arrayrulecolor{Gray}\hline
  	97$^a  $    & 1.168 & $^{+0.482}_{-0.406}$ & 3386.0 & $^{+72.5  }_{-72.5 }$ & 0.139 & $^{+0.062 }_{-0.058}$ & 0.296 & $^{+0.034 }_{-0.031}$ & 2.977  & $^{+6.142  }_{-1.566}$  & CTTS & TDC      & ASCC18        \\ \arrayrulecolor{Gray}\hline
  	98$^a  $    & 0.386 & $^{+0.589}_{-0.487}$ & 3386.0 & $^{+87.0  }_{-87.0 }$ & 0.13  & $^{+0.077 }_{-0.073}$ & 0.297 & $^{+0.047 }_{-0.051}$ & 3.271  & $^{+11.597 }_{-1.978}$  & WTTS & ClassIII & Outside       \\ \arrayrulecolor{Gray}\hline
  	107$^a $    & 0.33  & $^{+0.406}_{-0.406}$ & 3342.5 & $^{+72.5  }_{-72.5 }$ & 0.156 & $^{+0.082 }_{-0.082}$ & 0.272 & $^{+0.028 }_{-0.058}$ & 2.208  & $^{+6.233  }_{-1.237}$  & CTTS & Evolved  & Outside       \\ \arrayrulecolor{Gray}\hline
  	108         & 0.0   & $^{+0.513}_{-0.0  }$ & 3342.5 & $^{+87.0  }_{-87.0 }$ & 0.113 & $^{+0.066 }_{-0.06 }$ & 0.278 & $^{+0.025 }_{-0.076}$ & 3.513  & $^{+11.273 }_{-2.028}$  & WTTS & ClassIII & Outside       \\ \arrayrulecolor{Gray}\hline
  	109         & 3.533 & $^{+0.939}_{-1.066}$ & 3342.5 & $^{+145.0 }_{-145.0}$ & 0.157 & $^{+0.118 }_{-0.118}$ & 0.271 & $^{+0.075 }_{-0.071}$ & 2.188  & $^{+12.64  }_{-1.683}$  & CTTS & ClassII  & 25Ori         \\ \arrayrulecolor{Gray}\hline
  	110         & 0.0   & $^{+1.046}_{-0.0  }$ & 3342.5 & $^{+159.5 }_{-159.5}$ & 0.096 & $^{+0.086 }_{-0.069}$ & 0.282 & $^{+0.088 }_{-0.084}$ & 4.537  & $^{+10.338 }_{-3.771}$  & WTTS & ClassIII & Outside       \\ \arrayrulecolor{Gray}\hline
  	111         & 0.0   & $^{+0.487}_{-0.0  }$ & 3313.5 & $^{+87.0  }_{-87.0 }$ & 0.093 & $^{+0.046 }_{-0.04 }$ & 0.263 & $^{+0.037 }_{-0.063}$ & 4.353  & $^{+8.988  }_{-2.617}$  & WTTS & ClassIII & ASCC20        \\ \arrayrulecolor{Gray}\hline
  	115$^a $    & 0.046 & $^{+0.62 }_{-0.924}$ & 3255.5 & $^{+101.5 }_{-101.5}$ & 0.042 & $^{+0.027 }_{-0.029}$ & 0.207 & $^{+0.088 }_{-0.028}$ & 8.578  & $^{+6.298  }_{-5.968}$  & CTTS & TDC      & Outside       \\ \arrayrulecolor{Gray}\hline
  	116         & 0.0   & $^{+0.619}_{-0.0  }$ & 3255.5 & $^{+101.5 }_{-101.5}$ & 0.13  & $^{+0.083 }_{-0.074}$ & 0.218 & $^{+0.054 }_{-0.039}$ & 2.19   & $^{+8.071  }_{-1.685}$  & WTTS & Evolved  & Outside       \\ \arrayrulecolor{Gray}\hline
  	117         & 0.0   & $^{+0.67 }_{-0.0  }$ & 3241.0 & $^{+101.5 }_{-101.5}$ & 0.122 & $^{+0.07  }_{-0.058}$ & 0.206 & $^{+0.062 }_{-0.04 }$ & 2.204  & $^{+5.67   }_{-1.606}$  & WTTS & TDC      & ASCC20        \\ \arrayrulecolor{Gray}\hline
  	118$^a $    & 0.162 & $^{+0.721}_{-0.924}$ & 3226.5 & $^{+101.5 }_{-101.5}$ & 0.076 & $^{+0.044 }_{-0.047}$ & 0.2   & $^{+0.07  }_{-0.049}$ & 3.528  & $^{+11.281 }_{-2.451}$  & WTTS & Evolved  & ASCC20        \\ \arrayrulecolor{Gray}\hline
  	119         & 0.188 & $^{+1.097}_{-1.401}$ & 3212.0 & $^{+159.5 }_{-154.5}$ & 0.029 & $^{+0.028 }_{-0.03 }$ & 0.185 & $^{+0.115 }_{-0.085}$ & 10.461 & $^{+---    }_{-8.376}$  & WTTS & ClassIII & Outside       \\ \arrayrulecolor{Gray}\hline
  	120         & 0.0   & $^{+0.822}_{-0.0  }$ & 3197.5 & $^{+101.5 }_{-99.5 }$ & 0.052 & $^{+0.032 }_{-0.026}$ & 0.192 & $^{+0.058 }_{-0.067}$ & 4.924  & $^{+9.879  }_{-3.435}$  & WTTS & TDC      & ASCC20        \\ \arrayrulecolor{Gray}\hline
  	121$^a $    & 0.391 & $^{+0.823}_{-0.904}$ & 3197.5 & $^{+101.5 }_{-99.5 }$ & 0.021 & $^{+0.013 }_{-0.014}$ & 0.173 & $^{+0.067 }_{-0.048}$ & 14.128 & $^{+0.742  }_{-9.579}$  & CTTS & ClassII  & ASCC20        \\ \arrayrulecolor{Gray}\hline
  	123$^a $    & 0.0   & $^{+0.741}_{-0.0  }$ & 3197.5 & $^{+87.0  }_{-86.0 }$ & 0.068 & $^{+0.038 }_{-0.03 }$ & 0.198 & $^{+0.042 }_{-0.06 }$ & 3.635  & $^{+7.512  }_{-2.437}$  & WTTS & ClassII  & ASCC18        \\ \arrayrulecolor{Gray}\hline
  	124         & 0.0   & $^{+1.147}_{-0.0  }$ & 3197.5 & $^{+159.5 }_{-153.5}$ & 0.049 & $^{+0.044 }_{-0.037}$ & 0.191 & $^{+0.099 }_{-0.091}$ & 5.272  & $^{+9.602  }_{-4.006}$  & CTTS & Evolved  & 25Ori         \\ \arrayrulecolor{Gray}\hline
  	125         & 0.0   & $^{+0.792}_{-0.0  }$ & 3183.0 & $^{+87.0  }_{-85.0 }$ & 0.079 & $^{+0.045 }_{-0.036}$ & 0.192 & $^{+0.038 }_{-0.066}$ & 2.755  & $^{+6.192  }_{-1.742}$  & WTTS & TDC      & ASCC20        \\ \arrayrulecolor{Gray}\hline
  	126         & 0.0   & $^{+0.792}_{-0.0  }$ & 3183.0 & $^{+87.0  }_{-85.0 }$ & 0.042 & $^{+0.024 }_{-0.019}$ & 0.182 & $^{+0.042 }_{-0.056}$ & 5.805  & $^{+8.994  }_{-3.891}$  & WTTS & TDC      & ASCC20        \\ \arrayrulecolor{Gray}\hline
  	127$^a $    & 0.0   & $^{+0.66 }_{-0.0  }$ & 3183.0 & $^{+72.5  }_{-71.5 }$ & 0.14  & $^{+0.069 }_{-0.056}$ & 0.183 & $^{+0.039 }_{-0.045}$ & 1.118  & $^{+2.595  }_{-0.612}$  & WTTS & ClassIII & ASCC18        \\ \arrayrulecolor{Gray}\hline
  	128         & 0.0   & $^{+0.66 }_{-0.0  }$ & 3183.0 & $^{+72.5  }_{-71.5 }$ & 0.025 & $^{+0.014 }_{-0.013}$ & 0.171 & $^{+0.036 }_{-0.033}$ & 10.55  & $^{+4.332  }_{-6.362}$  & WTTS & ClassII  & Outside       \\ \arrayrulecolor{Gray}\hline
  	129$^a $    & 0.619 & $^{+0.66 }_{-0.64 }$ & 3168.5 & $^{+72.5  }_{-70.5 }$ & 0.09  & $^{+0.044 }_{-0.044}$ & 0.178 & $^{+0.028 }_{-0.053}$ & 1.966  & $^{+5.167  }_{-1.028}$  & CTTS & Evolved  & ASCC20        \\ \arrayrulecolor{Gray}\hline
  	130$^a $    & 0.0   & $^{+0.66 }_{-0.0  }$ & 3168.5 & $^{+72.5  }_{-70.5 }$ & 0.052 & $^{+0.026 }_{-0.021}$ & 0.18  & $^{+0.021 }_{-0.055}$ & 4.194  & $^{+7.27   }_{-2.586}$  & WTTS & TDC      & ASCC18        \\ \arrayrulecolor{Gray}\hline
  	131         & 0.0   & $^{+0.66 }_{-0.0  }$ & 3168.5 & $^{+72.5  }_{-70.5 }$ & 0.08  & $^{+0.04  }_{-0.033}$ & 0.18  & $^{+0.026 }_{-0.055}$ & 2.371  & $^{+4.579  }_{-1.325}$  & WTTS & ClassIII & ASCC18        \\ \arrayrulecolor{Gray}\hline
  	132$^a $    & 0.0   & $^{+0.528}_{-0.0  }$ & 3168.5 & $^{+58.0  }_{-57.0 }$ & 0.215 & $^{+0.112 }_{-0.1  }$ & 0.181 & $^{+0.019 }_{-0.043}$ & 0.623  & $^{+1.55   }_{-0.116}$  & WTTS & TDC      & Outside       \\ \arrayrulecolor{Gray}\hline
  	133         & 0.0   & $^{+0.66 }_{-0.0  }$ & 3168.5 & $^{+72.5  }_{-70.5 }$ & 0.087 & $^{+0.05  }_{-0.044}$ & 0.179 & $^{+0.028 }_{-0.053}$ & 2.075  & $^{+5.654  }_{-1.17 }$  & WTTS & ClassIII & Outside       \\ \arrayrulecolor{Gray}\hline
  	134         & 0.0   & $^{+0.528}_{-0.0  }$ & 3154.0 & $^{+58.0  }_{-56.0 }$ & 0.03  & $^{+0.013 }_{-0.011}$ & 0.163 & $^{+0.031 }_{-0.037}$ & 7.27   & $^{+7.53   }_{-4.032}$  & CTTS & TDC      & ASCC20        \\ \arrayrulecolor{Gray}\hline
  	135         & 0.0   & $^{+0.66 }_{-0.0  }$ & 3154.0 & $^{+72.5  }_{-69.5 }$ & 0.061 & $^{+0.044 }_{-0.036}$ & 0.173 & $^{+0.027 }_{-0.058}$ & 3.113  & $^{+10.263 }_{-1.943}$  & CTTS & Evolved  & ASCC18/ASCC20 \\ \arrayrulecolor{Gray}\hline
  	137         & 0.0   & $^{+0.66 }_{-0.0  }$ & 3154.0 & $^{+72.5  }_{-69.5 }$ & 0.06  & $^{+0.035 }_{-0.031}$ & 0.173 & $^{+0.027 }_{-0.058}$ & 3.222  & $^{+8.17   }_{-1.924}$  & WTTS & TDC      & Outside       \\ \arrayrulecolor{Gray}\hline
  	139         & 0.005 & $^{+1.137}_{-1.117}$ & 3154.0 & $^{+130.5 }_{-123.5}$ & 0.03  & $^{+0.026 }_{-0.026}$ & 0.163 & $^{+0.071 }_{-0.064}$ & 7.27   & $^{+7.593  }_{-5.211}$  & WTTS & TDC      & Outside       \\ \arrayrulecolor{Gray}\hline
  	141         & 0.0   & $^{+0.792}_{-0.0  }$ & 3125.0 & $^{+87.0  }_{-81.0 }$ & 0.037 & $^{+0.022 }_{-0.019}$ & 0.152 & $^{+0.048 }_{-0.052}$ & 4.762  & $^{+10.03  }_{-2.779}$  & WTTS & TDC      & ASCC18        \\ \arrayrulecolor{Gray}\hline
  	143         & 0.112 & $^{+1.564}_{-1.873}$ & 3125.0 & $^{+188.5 }_{-168.0}$ & 0.027 & $^{+0.035 }_{-0.037}$ & 0.149 & $^{+0.11  }_{-0.069}$ & 6.993  & $^{+---    }_{-5.394}$  & WTTS & ClassII  & Outside       \\ \arrayrulecolor{Gray}\hline
  	144         & 0.33  & $^{+1.401}_{-1.437}$ & 3125.0 & $^{+159.5 }_{-146.0}$ & 0.026 & $^{+0.029 }_{-0.029}$ & 0.148 & $^{+0.086 }_{-0.066}$ & 7.309  & $^{+---    }_{-5.331}$  & WTTS & ClassII  & Outside       \\ \arrayrulecolor{Gray}\hline
  	146         & 0.0   & $^{+0.782}_{-0.0  }$ & 3111.5 & $^{+86.0  }_{-81.0 }$ & 0.021 & $^{+0.018 }_{-0.016}$ & 0.139 & $^{+0.048 }_{-0.04 }$ & 8.428  & $^{+6.447  }_{-5.537}$  & WTTS & TDC      & 25Ori/ASCC20  \\ \arrayrulecolor{Gray}\hline
  	147         & 0.0   & $^{+0.782}_{-0.0  }$ & 3111.5 & $^{+86.0  }_{-81.0 }$ & 0.041 & $^{+0.034 }_{-0.029}$ & 0.146 & $^{+0.054 }_{-0.047}$ & 3.998  & $^{+10.89  }_{-2.444}$  & WTTS & Evolved  & 25Ori/ASCC18  \\ \arrayrulecolor{Gray}\hline
  	148         & 0.0   & $^{+0.914}_{-0.0  }$ & 3111.5 & $^{+100.5 }_{-94.5 }$ & 0.039 & $^{+0.026 }_{-0.022}$ & 0.146 & $^{+0.054 }_{-0.056}$ & 4.187  & $^{+10.589 }_{-2.437}$  & WTTS & Evolved  & 25Ori         \\ \arrayrulecolor{Gray}\hline
  	149         & 0.0   & $^{+0.914}_{-0.0  }$ & 3111.5 & $^{+100.5 }_{-94.5 }$ & 0.075 & $^{+0.056 }_{-0.049}$ & 0.145 & $^{+0.055 }_{-0.045}$ & 1.714  & $^{+10.208 }_{-1.208}$  & WTTS & ClassIII & Outside       \\ \arrayrulecolor{Gray}\hline
  	150         & 0.0   & $^{+0.761}_{-0.0  }$ & 3084.5 & $^{+84.0  }_{-81.0 }$ & 0.195 & $^{+0.12  }_{-0.103}$ & 0.162 & $^{+0.031 }_{-0.066}$ & 0.506  & $^{+1.392  }_{----  }$  & WTTS & ClassIII & ASCC20        \\ \arrayrulecolor{Gray}\hline
  	151         & 0.0   & $^{+0.63 }_{-0.0  }$ & 3084.5 & $^{+69.5  }_{-67.5 }$ & 0.136 & $^{+0.072 }_{-0.063}$ & 0.15  & $^{+0.027 }_{-0.056}$ & 0.842  & $^{+1.545  }_{-0.334}$  & WTTS & ClassIII & ASCC20        \\ \arrayrulecolor{Gray}\hline
  	152         & 0.0   & $^{+0.894}_{-0.0  }$ & 3084.5 & $^{+98.5  }_{-94.5 }$ & 0.027 & $^{+0.019 }_{-0.016}$ & 0.127 & $^{+0.057 }_{-0.036}$ & 5.486  & $^{+9.401  }_{-3.097}$  & CTTS & ClassII  & 25Ori         \\ \arrayrulecolor{Gray}\hline
  	153         & 0.0   & $^{+0.762}_{-0.0  }$ & 3084.5 & $^{+84.0  }_{-81.0 }$ & 0.047 & $^{+0.028 }_{-0.024}$ & 0.128 & $^{+0.053 }_{-0.032}$ & 2.666  & $^{+8.041  }_{-1.183}$  & CTTS & Evolved  & 25Ori         \\ \arrayrulecolor{Gray}\hline
  	154         & 0.0   & $^{+0.762}_{-0.0  }$ & 3084.5 & $^{+84.0  }_{-81.0 }$ & 0.057 & $^{+0.036 }_{-0.031}$ & 0.132 & $^{+0.046 }_{-0.036}$ & 2.18   & $^{+7.117  }_{-1.253}$  & WTTS & TDC      & ASCC18        \\ \arrayrulecolor{Gray}\hline
  	155         & 0.0   & $^{+0.894}_{-0.0  }$ & 3084.5 & $^{+98.5  }_{-94.5 }$ & 0.014 & $^{+0.01  }_{-0.008}$ & 0.119 & $^{+0.051 }_{-0.027}$ & 11.845 & $^{+3.031  }_{-7.036}$  & WTTS & ClassIII & 25Ori         \\ \arrayrulecolor{Gray}\hline
  	156$^a $    & 0.0   & $^{+1.025}_{-0.0  }$ & 3084.5 & $^{+113.0 }_{-105.5}$ & 0.038 & $^{+0.03  }_{-0.025}$ & 0.127 & $^{+0.072 }_{-0.046}$ & 3.463  & $^{+11.422 }_{-1.926}$  & WTTS & ClassII  & 25Ori         \\ \arrayrulecolor{Gray}\hline
  	157         & 0.0   & $^{+1.148}_{-0.0  }$ & 3071.0 & $^{+126.5 }_{-114.0}$ & 0.026 & $^{+0.026 }_{-0.021}$ & 0.12  & $^{+0.072 }_{-0.042}$ & 5.335  & $^{+9.537  }_{-3.422}$  & WTTS & TDC      & Outside       \\ \arrayrulecolor{Gray}\hline
  	158         & 0.0   & $^{+1.777}_{-0.0  }$ & 3030.5 & $^{+196.0 }_{-167.5}$ & 0.019 & $^{+0.032 }_{-0.023}$ & 0.098 & $^{+0.102 }_{-0.026}$ & 6.334  & $^{+---    }_{-5.676}$  & CTTS & ClassII  & Outside       \\ \arrayrulecolor{black}\hline
\enddata
\tablenotetext{a}{$>99\%$ probable variable star according to the CVSO study \citep{Briceno2005,Mateu2012}.}
\tablecomments{Outside location label indicates the members not belonging to any stellar group defined by \citet{Kharchenko2013}.}
\end{deluxetable}

\subsubsection{Effective Temperatures and Bolometric Luminosities}
\label{subsubsec:HR}

We estimated the effective temperatures of the confirmed members by interpolating their spectral types into the empirical relations from \citet{Luhman1999}.

\begin{figure*}[ht!]
	\includegraphics[width=1.\textwidth]{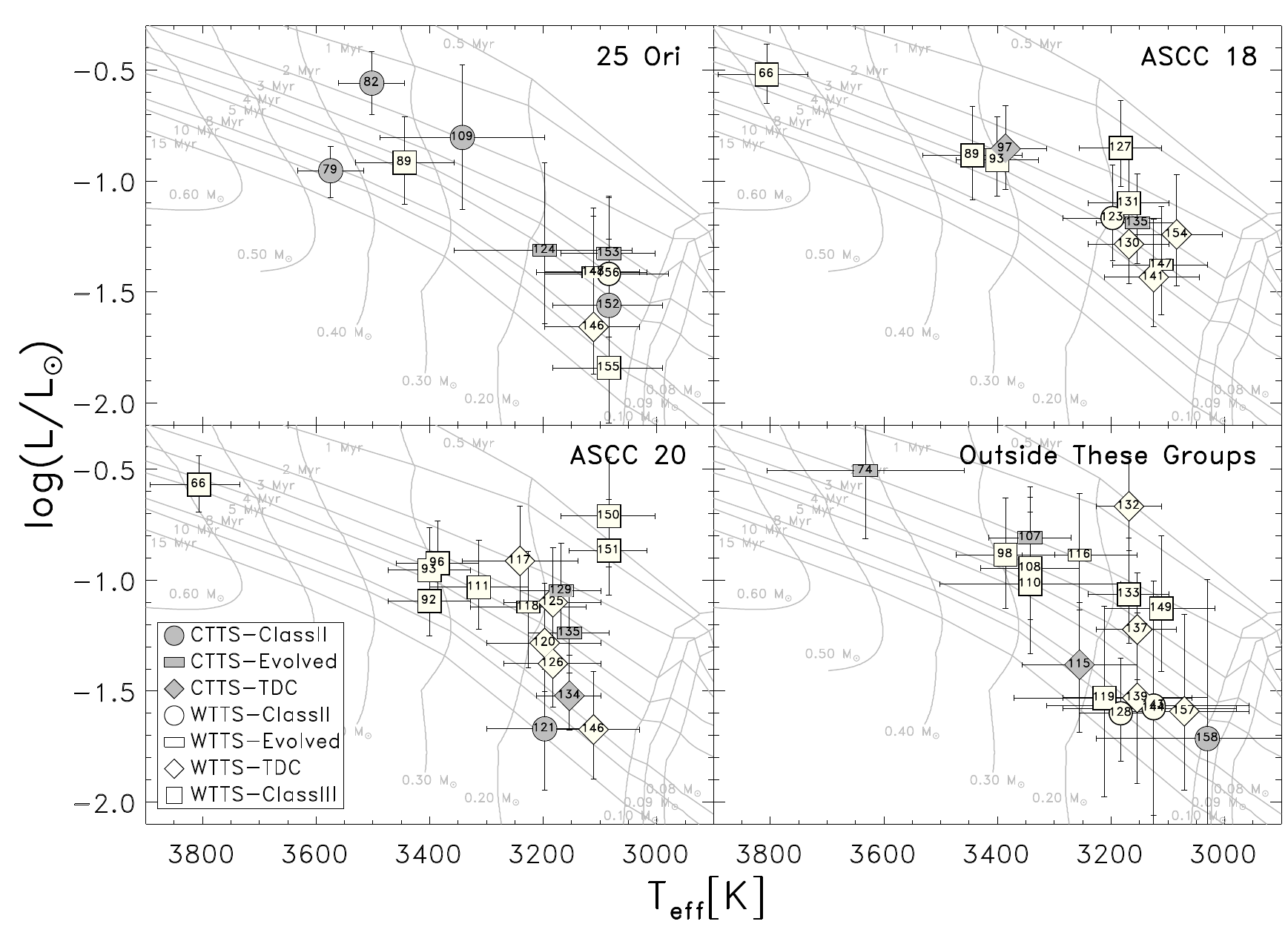}
	\caption{H-R diagrams of the confirmed members inside the labeled stellar groups or outside them, according to \citet{Kharchenko2013}. The gray curves represent the PMS \citet{Baraffe2015} models. The members overlapping two stellar groups (see Figure \ref{fig:sky}) appear in both H-R diagrams.
	\label{fig:H-R}}
\end{figure*}

To compute the bolometric luminosities of these members, we used newly available Gaia data \citep[DR1; ][]{Gaia2016} to establish distances for the 25 Ori, ASCC 18, and ASCC 20 stellar groups, where the confirmed members are located. In Table \ref{tab:distances} we summarize the previous distances to these groups \citep{Kharchenko2005,Briceno2005,Briceno2007,Kharchenko2013,Downes2014}, as well as our own estimates from the Gaia parallaxes for the higher probability \citet{Kharchenko2005} members of these groups. The Gaia-based distance estimates for the 25 Ori, ASCC 18, and ASCC 20 stellar groups are, within the uncertainties, essentially identical of 338$\pm$66 pc. We then used individual distance estimates derived from Gaia parallaxes to calculate the absolute $I$ magnitudes (M$_\textrm{\scriptsize {I}}$) for the confirmed members projected inside these stellar groups. For members located outside these groups, we assumed the mean Gaia distance. Then we used the bolometric correction from \citet{KH1995} to obtain the bolometric luminosity, assuming M$_{\textrm{\scriptsize{bol}}_\odot}=4.755$ \citep{Mamajek2012}. In Figure \ref{fig:H-R} we show the locations of the confirmed members in the H-R diagrams according to the stellar group where they lie or if they are outside (see Figure \ref{fig:sky}) and in Table \ref{tab:parameters} are listed the effective temperatures and bolometric luminosities we estimated for the confirmed members.

%\floattable
\begin{deluxetable}{lccc}
\tabletypesize{\scriptsize}
\tablecaption{Distances of the stellar groups partially covered by the 25 Ori BOSS plate.
\label{tab:distances}}
\tablewidth{0pt}
\tablehead{
\colhead{Reference} & \multicolumn{3}{c}{Distance} \\
	            & \multicolumn{3}{c}{(pc)} \\
\cline{2-4}
 	            & 25 Ori & ASCC 18 & ASCC 20
}
\colnumbers
\startdata
	\citet{Kharchenko2005}          & 460            & 500            & 450 \\ \arrayrulecolor{Gray}\hline
	\citet{Kharchenko2013}          & 397            & 313            & 394 \\ \arrayrulecolor{Gray}\hline
	\citet{Briceno2005,Briceno2007} & 330            & ...            & ... \\ \arrayrulecolor{Gray}\hline
	\citet{Downes2014}              & 360            & ...            & ... \\ \arrayrulecolor{Gray}\hline
	\citet{Gaia2016}                & 336$\pm$30$^a$ & 349$\pm$44$^b$ & 330$\pm$39$^c$ \\  \arrayrulecolor{black}
\enddata
\tablenotetext{a}{for 17 high-probability members.}
\tablenotetext{b}{for 7 high-probability members.}
\tablenotetext{c}{for 15 high-probability members.}
\end{deluxetable}

\subsubsection{Masses and Ages}

As shown in Figure \ref{fig:H-R}, all members are scattered, within the uncertainties, throughout the 1-10 Myr isochrones and 0.1-0.6 M$_\odot$ evolutionary tracks, adopting the PMS models of \citet{Baraffe2015}.

To better estimate the masses and ages of the confirmed members, we interpolated their effective temperatures and bolometric luminosities into the \citet{Baraffe2015} models. When a confirmed member overlaps two stellar groups we adopted its mean bolometric luminosity. In Table \ref{tab:parameters} we show the mass and age we obtained for each confirmed member. The resulting masses for the 53 confirmed members are in a range from 0.10 M$_\odot$ to 0.58 M$_\odot$, with $64\%$ of the members having masses lower than 0.20 M$_\odot$. This implies that in this mass range we increased by a $\approx50\%$ the number of confirmed LMSs in the region covered by the BOSS plate, and by a $\approx$20\% the number of LMSs in the estimated area of 25 Ori \citep[1$^\circ$ radius; ][]{Briceno2005,Briceno2007}.

The ages we calculated for the confirmed members are roughly twice younger than those found with similar methods in previous studies for the stellar groups where they are located \citep{Kharchenko2005,Briceno2005,Briceno2007,Kharchenko2013,Downes2014}. This is due to the target selection bias in the 25 Ori BOSS spectra (see Section \ref{spectroscopy} and Figure \ref{fig:CCD_bias}).

For all the confirmed members we estimated the uncertainties in the derived values of the main physical parameters (extinction, effective temperature, bolometric luminosity, mass, and age) by considering the following factors: the error propagation that applies to each step described in this section and the errors associated to the interpolations. In Table \ref{tab:parameters} we show the resulting uncertainties for the extinction, effective temperature, bolometric luminosity, mass, and age values for the confirmed members. We also indicate to which stellar group they belong.

\subsection{T Tauri Star Classification} 
\label{subsec:TTSclass}

\begin{figure}%[H]
	\includegraphics[width=.5\textwidth]{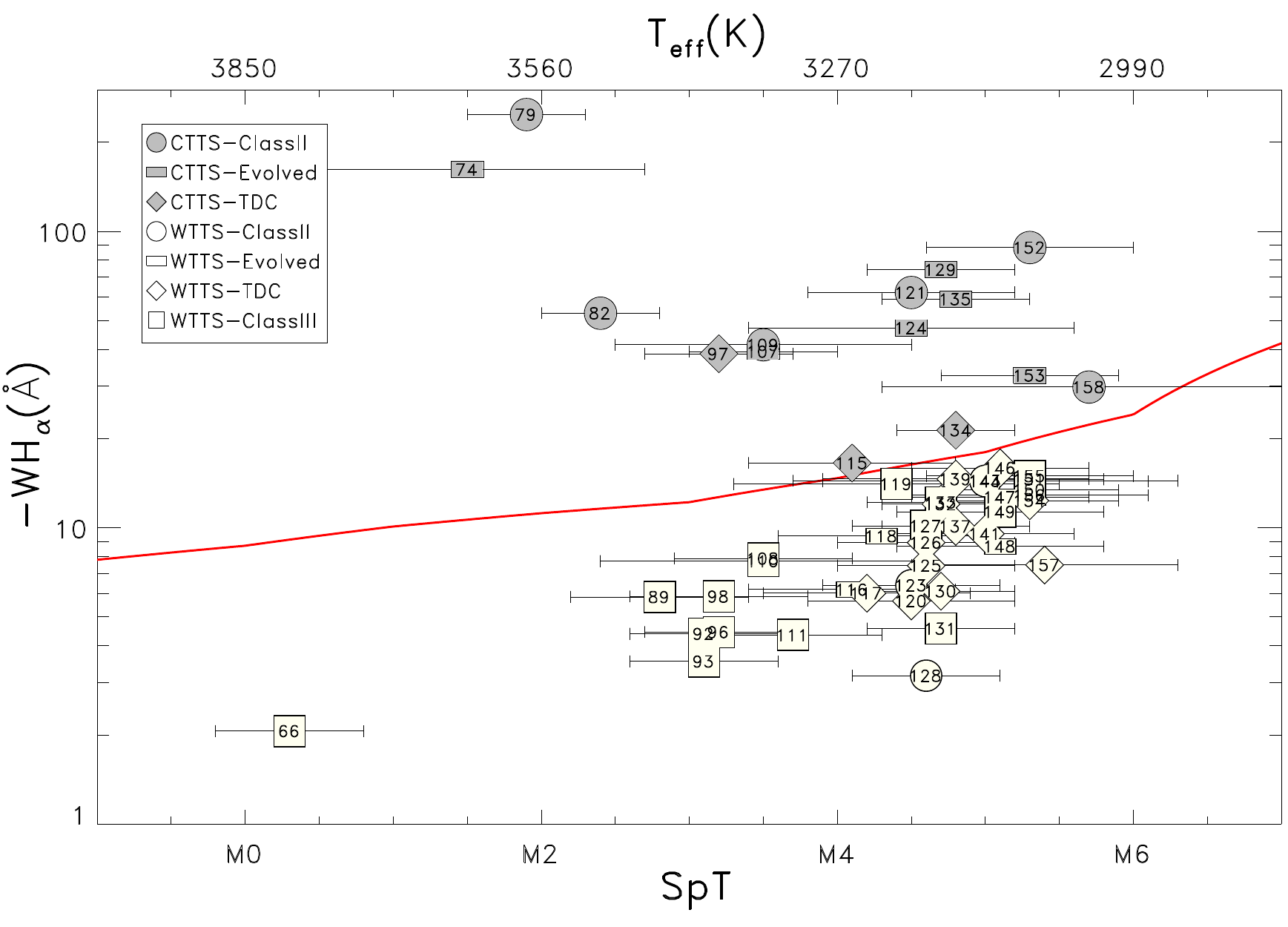}
	\caption{Relation between the H$_\alpha$ equivalent widths and spectral types for the confirmed members. The red solid curve indicates the saturation limit from \citet{Barrado2003}, which allow us to separate WTTSs from CTTSs. The upper axis shows the effective temperature corresponding to the spectral types \citep{Luhman2003b}.}
	\label{fig:WHavsSpT}
\end{figure}

The BOSS low-resolution spectra allowed us to measure the H$_\alpha$ equivalent width, which we used, together with the spectral types, to classify the confirmed members as either accreting young LMSs (classical T Tauri stars; CTTSs) or non-accreting young LMSs (weak T Tauri stars; WTTSs). To define the limit between both types of T Tauri stars (TTSs), we adopted the empirical saturation criterion of \citet{Barrado2003}, in which stars with H$_\alpha$ emission above this limit are classified as CTTSs, and otherwise as WTTSs. In Figure \ref{fig:WHavsSpT} we show the TTS classification scheme and in Table \ref{tab:parameters} we show the resulting classification for the whole sample of confirmed members.

We confirmed a total of 15 CTTSs and 38 WTTSs among the 53 members in the BOSS sample. This corresponds to a very high fraction of CTTSs to WTTSs compared to the values of 5.6\% and 3.8$\pm0.4$\% found by \citet{Briceno2007} and \citet{Downes2014}, respectively, which is due to the target selection bias toward sources with IR excesses (see Section \ref{spectroscopy} and Figure \ref{fig:CCD_bias}).

\subsection{Spectral Energy Distributions}
\label{subsec:SED}

The circumstellar disks of the YSOs were classified according to their IR excess emissions at $\lambda>2\ \mu$m. Objects having a flat or decreasing IR spectral energy distribution (SED) are considered Class II, while Class III objects have little or no near-IR excess \citep{LW1984,Lada1987}. An intermediate phase between the Class II and Class III objects contains the so-called ``transitional disk systems" that present a decreasing SED slope in the near-IR that rises again in the mid-IR. Finally, the ``evolved disk systems" show a monotonically decreasing IR SED \citep[e.g.,][]{Hernandez2007b}.

To classify the IR excesses of the confirmed members according to this scheme, we constructed their SEDs using the Virtual Observatory SED Analyzer (VOSA) tool \citep{Bayo2008} and the photometric catalogs described in Sections \ref{subsec:Optphot} and \ref{subsec:IRphot}, and listed in Table \ref{tab:photometry}. We have a minimum of 10 and a maximum of 24 photometric bands for each confirmed member, covering a wavelength range from 0.36 $\mu$m to 22 $\mu$m.

The SEDs were dereddened using the visual extinction we estimated in Section \ref{subsubsec:extinction} and assuming the extinction law reported by \citet{Fitzpatrick1999} and subsequently improved by \citet{Indebetouw2005}. To determine the corresponding IR excesses we proceeded iteratively as follows: First, we fitted the SEDs to the PMS LMS models from \cite{Baraffe2015}, restricting the effective temperature range to the one obtained in Section \ref{subsubsec:HR}. During this iteration we only considered the photometric bands where the IR excesses are not expected to occur ($\lambda < 2\ \mu$m).

From the resulting fitted SEDs, VOSA automatically detects which bands present IR excesses by using an improved algorithm from that of \citet{Lada2006}, which measures the slope of the IR points in the log($\lambda F_\lambda$) vs log($\lambda$) space. Basically, when the slope becomes greater than -2.56, the IR excesses are determined.

\begin{figure*}[ht!]
	\centering
	\includegraphics[width=1.0\textwidth]{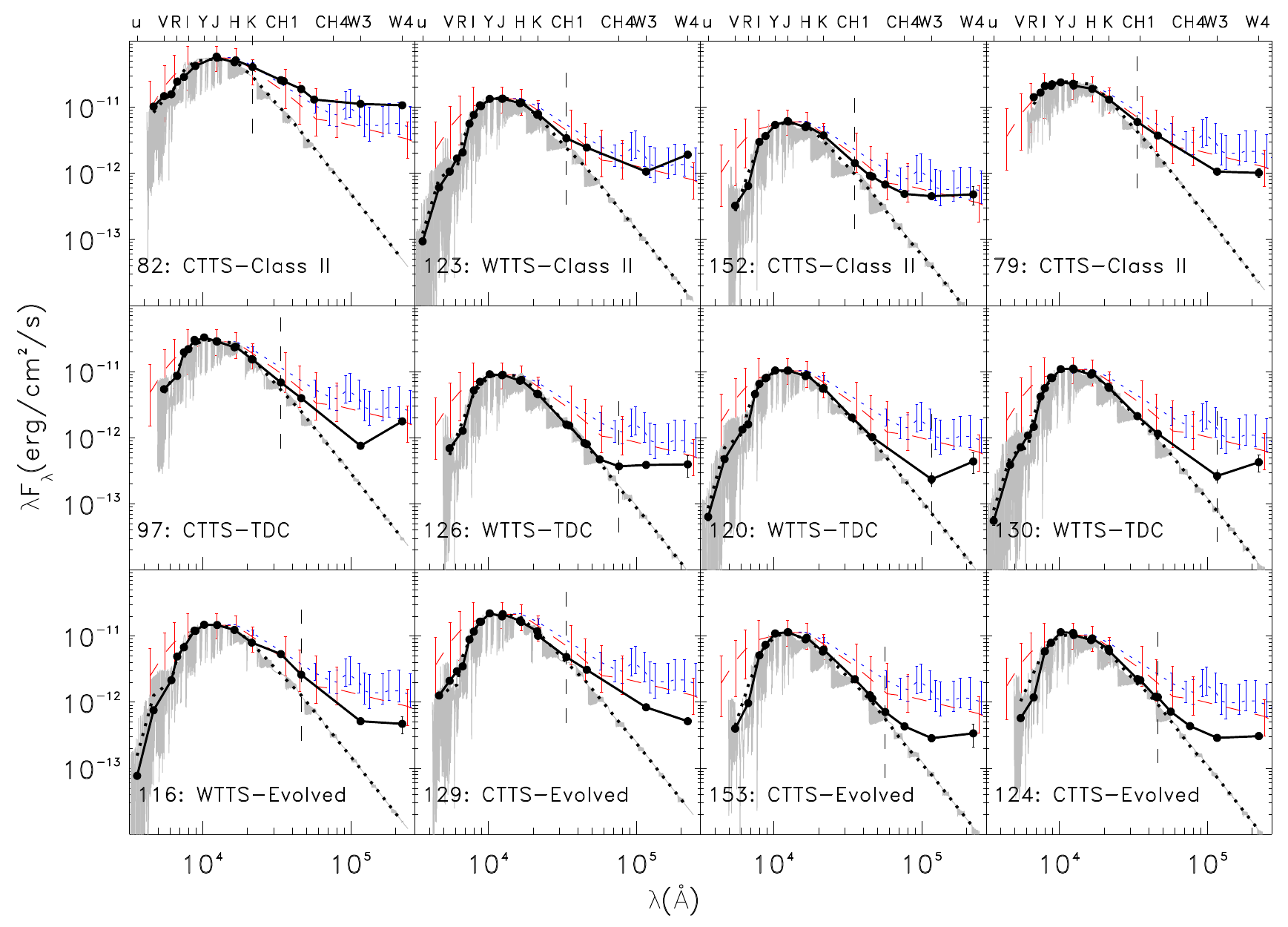}
	\caption{Dereddened SEDs for a sample of the confirmed members (black points and black solid curves). The gray spectra correspond to the best PMS LMS model from \citet{Baraffe2015} fitted to the dereddened data bluer than the point were the IR excesses start (vertical black dashed line). The black dotted curves show the fitted \citet{Baraffe2015} model spectra in a lower resolution. The red dashed curves and the blue dotted ones indicate, respectively, the median SEDs of Class II disks of the $\sigma$ Orionis cluster \citep{Hernandez2007a} and the Taurus star-forming region \citep{Furlan2006}, normalized to the dereddened $J$-band flux of each member. The vertical red and blue solid lines represent the upper and lower quartiles for these median SEDs. Each SED has a label with the member ID and its TTS and disk classifications, as explained in Sections \ref{subsec:TTSclass} and \ref{subsec:SED}, respectively. The photometric errors are included but most of them are smaller than the corresponding symbols. All the SEDs of the confirmed members are available in the electronic version of this publication.}
	\label{fig:SEDs}
\end{figure*}

Then, a second fit to the \cite{Baraffe2015} models was performed, this time excluding those photometric bands showing IR excesses. In this way, we avoided false IR excess detections during the first iteration and maximized the number of photometric bands used for fitting the photospheres. The number of photometric bands used during the second iteration ran from 10 to 23 (except for the members 74 and 109 with 7 and 8 fitted bands, respectively). In Figure \ref{fig:SEDs} we show the resulting SEDs for a selection of the confirmed members. The bolometric luminosities for the confirmed members corresponding to the total flux of the best \citet{Baraffe2015} model fit are in agreement, within the uncertainties, with those obtained using the $I$ band, as explained in Section \ref{subsubsec:HR}.

We classified the members as Class II if their IR SEDs resemble the median SEDs of Class II disks of the $\sigma$ Orionis cluster \citep{Hernandez2007a} and the Taurus star-forming region \citep{Furlan2006}. The members showing lower IR excesses were considered evolved systems, while the members having IR SEDs consistent with the photospheric \citet{Baraffe2015} model fit were classified as Class III. The members showing a near-IR SED consistent with evolved systems or Class III objects, but having an unexpected strong excess at 22 $\mu$m were considered as transitional disk candidates (TDC).

\begin{figure*}[ht!]
	\centering
	\includegraphics[width=0.8\textwidth]{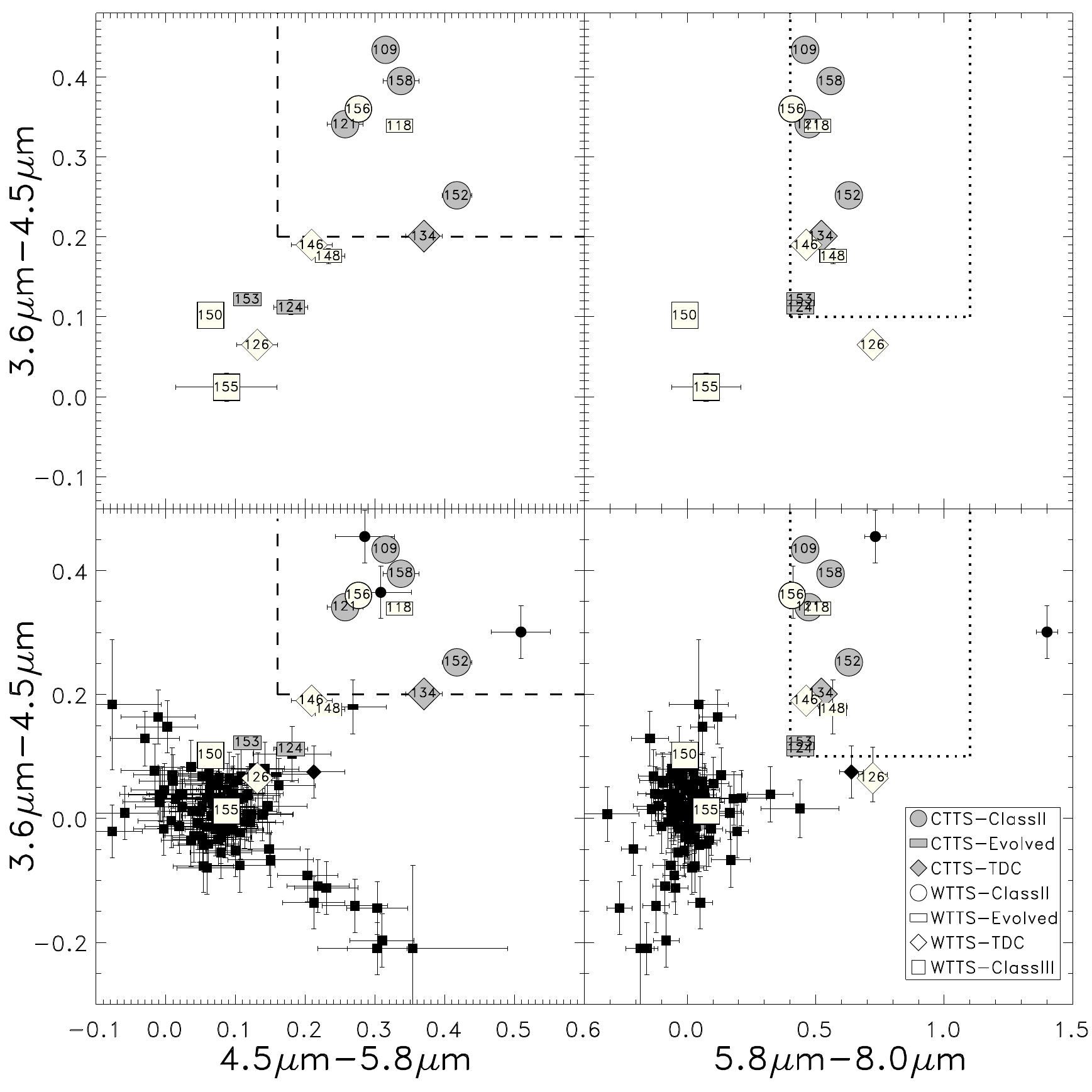}
	\caption{IRAC color-color diagrams for the confirmed members from this work (top panels) and including those from \citet{Hernandez2007b} (bottom panels). The small black filled circles, diamonds, horizontal bars and square represent YSOs with Class II, transitional, evolved or Class III disks from \citet{Hernandez2007b}. The dashed lines delimit the region where M type objects with disk are expected, from the study of \citet{Luhman2005}, and the dotted lines show the CTTS locus from \citet{Hartmann2005}.}
	\label{fig:irac}
\end{figure*}

For the 14 members having available photometry in the [3.6] and [8.0] bands from IRAC, the slope in the [3.6]-[8.0] color ($\alpha$, in the log[$\lambda F_\lambda$] vs log[$\lambda$] space) was analyzed to improve the disk classification as follows \citep{Lada2006}: Class II objects have slopes of $-1.8<\alpha<0$; evolved or ``anemic" disk systems \citep{Hernandez2007a} have $-2.56\le\alpha<-1.8$ slopes; Class III objects have $\alpha<-2.56$. In Figure \ref{fig:irac} we show the locations of the members in the IRAC color-color diagrams. All the CTTSs fall inside the CTTS locus defined by \citet{Hartmann2005}, but four objects classified as WTTSs also fall inside this region (two having evolved disks, one is a TDC and the other one is bearing a Class II disk). All the members located in the IR excess region defined by \citet{Luhman2005} have Class II disks and only one of them has an evolved disk. The rest of the evolved systems and TDCs are located in the region between the Class II and Class III objects, as expected from the \citet{Hernandez2007b} sample. 

Of the 53 confirmed members we classified: a) 11 Class II objects, with SEDs consistent with the $\sigma$ Orionis cluster and Taurus star-forming region median SEDs; b) 10 evolved disks, showing decreasing IR excesses but smaller than the aforementioned medians; c) 15 TDCs, having a sudden increase in their IR excesses at $22\ \mu$m; d) 17 Class III, with no detectable IR excesses. In Table \ref{tab:parameters} we list the final disk type classification for the LMS members. For the sources showing IR excesses, those start at the WISE 3.4 $\mu$m band (for $\approx 42$\% of them) or longer wavelengths. Only for one member (member 82), the IR excess starts in the $K$ band.

Considering both TTS and disk classifications for the 53 confirmed members: 17 out of the 38 WTTSs have disks of Class III, 12 are TDCs, 4 are evolved systems and 5 have Class II disks. All the 15 CTTSs show IR excesses, with 6 having Class II disks, 6 evolved systems and 3 TDCs.

\section{Peculiar Objects}
\label{sec:singular}

\subsection{Variable Members}

Variability is an important effect than can be present in the member sample. It could modify their locations in the color-magnitude diagrams and affect the determination of physical parameters such as extinction, bolometric luminosity, mass, and age. We expect that the variability in the $I$ band should not have important effects in our confirmed member sample because we are using the CDSO photometric catalog, which lists mean magnitudes of multi-epoch observations with temporal spacing of about 4 yr (see Section \ref{subsec:Optphot}). However, we are also working with multi-epoch VISTA photometry, which has temporal spacing of only 14 nights, where variability can be present. About $34\%$ of the confirmed members have $>99\%$ probability of being variable stars according to the CVSO catalog \citep{Briceno2005,Mateu2012}.

In the $I$ vs $I-J$ color-magnitude diagram (see Figure \ref{fig:IvsI-J}), members 110, 116, 125, and 131 fall outside the region defined by the YSO candidates. None of these members is listed as a high-probability variable stars in the CVSO catalog. However, they show the greatest $J_\textrm{\tiny {VISTA}}-J_\textrm{\tiny {2MASS}}$ residuals (together with the high-probability variable stars 66, 74, and 118), with values of: 0.714, 0.357, 0.163, and 0.415 mag for members 110, 116, 125, and 131, respectively, which are, within the uncertainties, significantly larger than those for the rest of the members. These $J$-band differences explain well the deviated positions only for members 110 and 131. Members 110, 116, and 125 have close sources in the SDSS or 2MASS images, which may be contaminating their photometries, causing their deviations in the $I$ vs $I-J$ diagram.

\subsection{High Extinction Members}
\label{sec:high_extinction}

Considering that the mean extinction toward 25 Ori is $\bar{A}_V\approx$0.28 mag \citep{Kharchenko2005, Briceno2005, Briceno2007, Downes2014}, members 74 and 109 present significantly higher extinction values of $A_V=4.33^{+0.51}_{-0.98}$ and $A_V=3.53^{+0.94}_{-1.01.}$ mag, respectively. These members are not high-probability variable stars in the CVSO catalog, although they have the largest $I-J$ colors in the sample (see Figure \ref{fig:IvsI-J}). Furthermore, members 74 and 109 were classified as CTTSs showing, respectively, an evolved disk system and a Class II disk. Additionally, the spectra of these two members show IR emissions more intense than those for the confirmed members with the same spectral types but with low extinction values. It may be that their high-extinction values are caused by dust in their disks, which are presented to us with an edge-on geometry. The positions of these members in the H-R diagram (see Figure \ref{fig:H-R}) are, within the uncertainties, consistent with most of the members. 

\subsection{Highly Luminous Members}

The deviant position of few members (132, 150, and 151) from the rest of the sample in the H-R diagrams (see Figure \ref{fig:H-R}) can be naturally explained by their effective temperature and bolometric luminosity uncertainties. Only the member 132 is a $>99\%$ probability variable star according to the CVSO catalog. Members 132 and 150 have a close companion in the SDSS or 2MASS images, which can be contaminating their photometries. The member 151 may be an isolated star without signals of variability, indicating that its position in the H-R diagram could be real.

\section{Discussion and Conclusions}
\label{sec:summary}

We determined the memberships of LMSs in the SDSS-III/BOSS spectra in 25 Ori and Orion OB1a on the basis of the presence of H$_\alpha$ emission and either LiI$\lambda$6708 or weak NaI$\lambda\lambda$8183, 8195 absorptions. We confirm 53 LMS members of 25 Ori or Orion OB1a, of which only three have been confirmed before by \citet{Downes2014}. These members are located in regions associated with at least three different stellar groups belonging to Orion OB1a \citep[25 Ori, ASCC 18, and ASCC 20; ][]{Kharchenko2005,Kharchenko2013}. The new LMS sample represents an increase of $\approx$50\% in the number of M0-M6 spectral type spectroscopically confirmed members in the area of the 25 Ori BOSS plate and a $\approx$20\% increase in the number of LMSs known inside the 25 Ori's estimated area \citep[1$^\circ$ radius; ][]{Briceno2005,Briceno2007}.

We did not confirm any K-type member in the 25 Ori BOSS plate on the basis of the H$_\alpha$ emission and LiI$\lambda$6708 absorption criteria. Furthermore, we found that the stars earlier than K-type are likely field stars after checking their position in the $I$ vs $I-J$ color-magnitude diagram and looking for their X-ray emission, IR excesses or variability.

Parallaxes for high-probability members from the \citet{Kharchenko2005} list are available from the Gaia DR1 catalog \citep{Gaia2016}. Using these parallaxes, we derived distances of 336$\pm$30, 349$\pm$44, and 330$\pm$39 pc for 25 Ori, ASCC 18, and ASCC 20, respectively. Within the uncertainties, these stellar groups are located at the same distance (338$\pm$66 pc), but our estimates are based on a small number of high-probability members (17, 7, and 15 for 25 Ori, ASCC 18, and ASCC 20, respectively). With the next Gaia release we will have parallaxes for many more high-probability members and even for confirmed sub-solar members. 

The mean extinction (excluding two outliers) we calculated toward the whole member sample is $\bar{A}_V$=0.14 mag. If we only consider the members inside the 25 Ori area, we obtained $\bar{A}_V$=0.21 mag, which is slightly lower than the one in previous studies \citep[0.27 mag, 0.28 mag, 0.29 mag, and 0.30 mag by ][]{Kharchenko2005, Briceno2005, Briceno2007, Downes2014}. This small difference may be caused by the fact that our confirmed members in Orion OB1a span towards the south-east of the 25 Orionis star (see Figure \ref{fig:sky}), where the \citet{Schlegel1998} extinction is even lower than in the area closer to the 25 Orionis star, as show in Figure 1 from \citet{Downes2014}. Members 74 and 109 have extinctions of $A_V=4.33^{+0.51}_{-0.98}$ and $A_V=3.53^{+0.94}_{-1.01.}$ mag, respectively, which are much higher than the mean. A likely explanation could be that these members present edge-on disks, similar to the BD 4 member from \citet{Downes2015}.

We constructed H-R diagrams for the confirmed members (see Figure \ref{fig:H-R}), assuming the distances determined from Gaia parallaxes (see Table \ref{tab:distances}). According to the PMS models from \citet{Baraffe2015}, the mass range covered by the members is from 0.10 M$_\odot$ to 0.58 M$_\odot$. We do not find a clear separation over the isochrones for the members located in the different stellar groups. The ages we estimated for the confirmed members are younger by a factor of $\sim 2$ than those for the stellar groups in which they lie \citep{Kharchenko2005,Briceno2005,Briceno2007,Kharchenko2013,Downes2014}. This is due to a bias in the target selection for the 25 Ori BOSS plate toward members with IR excesses (see Section \ref{spectroscopy}). This bias is clear in Figure \ref{fig:CCD_bias}, where most of the BOSS targets have $K$-W3 colors redder than those expected from previously confirmed members. 

Following the empirical saturation criterion by \citet{Barrado2003} for the TTSs classification of the confirmed members, we found 38 WTTSs and 15 CTTSs. This number of CTTSs is very high, considering that the fraction of CTTSs in 25 Ori has a mean value of 4.7\% \citep{Briceno2007,Downes2014}, which is due to the bias in the selection of targets for the 25 Ori BOSS plate.

%We classified the 53 confirmed members as either WTTSs or CTTSs, following the empirical saturation criterion by \citet{Barrado2003}. We found a CTTS to WTTS fraction of about 40\%, which is very high compared with previous estimates of 5.6\% and 3.8$\pm0.4$\% by \citet{Briceno2007} and \citet{Downes2014}, respectively. Again, this high fraction of CTTSs is due to the bias in the BOSS target selection.

We constructed the SEDs of the TTSs and fitted the photospheric \citet{Baraffe2015} models in order to detect the IR excesses and classify their disks. We found: 11 Class II disks, with SEDs consistent with the median SEDs of Class II disks of the $\sigma$ Orionis cluster \citep{Hernandez2007a} and the Taurus star-forming region \citep{Furlan2006}; 10 evolved disks, with falling IR SEDs showing excesses smaller than the medians SEDs; 15 TDCs, with falling near-IR SEDs with a sudden increase in the mid-IR; and 17 Class III disks, with SEDs with no detectable IR excesses, consistent with the photospheric \citet{Baraffe2015} models. For the members showing IR excesses, these start at wavelength longer that WISE 3.4 $\mu$m (only for the member 82 these start in the $K$ band), which assure that the masses we assigned to the TTSs, working with the $I$ and $J$ bands, are not affected by the IR excesses.

The 34\% of the confirmed members are $>99\%$ probability variable stars in the CVSO catalog \citep{Briceno2005,Mateu2012}. This effect, together with close sources to the members in the SDSS and 2MASS images, explained well most of the deviated members in the $I$ vs $I-J$ color-magnitude diagram (see Figure \ref{fig:IvsI-J}) and H-R diagrams (see Figure \ref{fig:H-R}). Only the position of the member 151 in the H-R diagrams, that appears younger that expected, seems to be real. Additional analysis are necessary to reveal the nature of this object.

\subsection{Chromospheric Activity}
Due to the bias in the target selection for the 25 Ori BOSS plate, many of the confirmed members exhibit very strong H$\alpha$ emission. This intense emission is due to strong chromospheric (magnetic) activity for the WTTSs and a combination of this phenomenon with ongoing accretion for the CTTSs. A number of recent studies have demonstrated that chromospheric activity in LMSs can alter their physical properties relative to the expectations of non-magnetic stellar models. In particular, strong activity appears to be able to inflate the stellar radius and to decrease the effective temperature \citep[e.g. ][]{Lopez-Morales2007,Morales2008}. Typical amounts of radius inflation and effective temperature suppression are $\sim$10\% and $\sim$5\%, respectively \citep{Lopez-Morales2007}.

For PMS LMSs, these effects can be quite important, causing the stars to appear to have lower masses and younger ages. For example, \citet{Stassun2012} developed empirical relations for the radius inflation and effective temperature suppression for a given amount of chromospheric H$\alpha$ luminosity. These relations predict that the effective temperature suppression and radius inflation roughly preserve the bolometric luminosity. In addition, \citet{Stassun2014} showed that the effect of effective temperature suppression on ensembles of young LMSs and BDs is to skew the inferred initial mass function strongly toward lower masses.  

In the case of the LMSs studied here, the individual ages we have determined for the entire sample are slightly younger than the mean estimated age of 25 Ori from previous studies. The combined effects of the effective temperature suppression from chromospheric activity as well as the bias in the target selection towards sources harboring disks, could explain such result. If so, then some of these stars could have higher masses and to be slightly older. This would have the effect of narrowing the age spread found here for these Orion OB1a groups. A detailed characterization of the mean ages of these regions is, however, beyond the scope of this work as it would require of a more robust sample of members.

\section*{ACKNOWLEDGEMENTS}

GS acknowledges support from a CONACYT/UNAM Mexico doctoral fellowship.

CRZ, JJD and GS acknowledge support from programs UNAM-DGAPA-PAPIIT IN116315 and IN108117, Mexico. MT also acknowledges grant No. IN-104316.

JJD acknowledges support from Secretar\'ia de Relaciones Exteriores del Gobierno de M\'exico.

% SDSS-III/BOSS
Funding for SDSS-III has been provided by the Alfred P. Sloan Foundation, the Participating Institutions, the National Science Foundation, and the U.S. Department of Energy Office of Science. The SDSS-III web site is \url{http://www.sdss3.org/}.

SDSS-III is managed by the Astrophysical Research Consortium for the Participating Institutions of the SDSS-III Collaboration including the University of Arizona, the Brazilian Participation Group, Brookhaven National Laboratory, Carnegie Mellon University, University of Florida, the French Participation Group, the German Participation Group, Harvard University, the Instituto de Astrofisica de Canarias, the Michigan State/Notre Dame/JINA Participation Group, Johns Hopkins University, Lawrence Berkeley National Laboratory, Max Planck Institute for Astrophysics, Max Planck Institute for Extraterrestrial Physics, New Mexico State University, New York University, Ohio State University, Pennsylvania State University, University of Portsmouth, Princeton University, the Spanish Participation Group, University of Tokyo, University of Utah, Vanderbilt University, University of Virginia, University of Washington, and Yale University.

% SDSS photometry
Funding for the SDSS and SDSS-II has been provided by the Alfred P. Sloan Foundation,
the Participating Institutions, the National Science Foundation, the U.S. Department
of Energy, the National Aeronautics and Space Administration, the Japanese Monbukagakusho,
the Max Planck Society, and the Higher Education Funding Council for England. The SDSS
Web Site is \url{http://www.sdss.org/}. The SDSS is managed by the Astrophysical Research
Consortium for the Participating Institutions. The Participating Institutions are the
American Museum of Natural History, Astrophysical Institute Potsdam, University of Basel,
University of Cambridge, Case Western Reserve University, University of Chicago,
Drexel University, Fermilab, the Institute for Advanced Study, the Japan Participation
Group, Johns Hopkins University, the Joint Institute for Nuclear Astrophysics, the
Kavli Institute for Particle Astrophysics and Cosmology, the Korean Scientist Group,
the Chinese Academy of Sciences (LAMOST), Los Alamos National Laboratory,
the Max-Planck-Institute for Astronomy (MPIA), the Max-Planck-Institute for
Astrophysics (MPA), New Mexico State University, Ohio State University, University
of Pittsburgh, University of Portsmouth, Princeton University, the United States Naval
Observatory, and the University of Washington.

% CDSO 
Based on observations obtained at the Llano del Hato National Astronomical Observatory of Venezuela, operated by the Centro de Investigaciones de Astronom{\'\i}a (CIDA) for the Ministerio del Poder Popular para Educaci\'on Universitaria, Ciencia y Tecnolog{\'\i}a.

%GAIA
This work has made use of data from the European Space Agency (ESA)
mission {\it Gaia} (\url{http://www.cosmos.esa.int/gaia}), processed by
the {\it Gaia} Data Processing and Analysis Consortium (DPAC,
\url{http://www.cosmos.esa.int/web/gaia/dpac/consortium}). Funding
for the DPAC has been provided by national institutions, in particular
the institutions participating in the {\it Gaia} Multilateral Agreement.

%2MASS
This publication makes use of data products from the Two Micron All Sky Survey, which is a joint project of the University of Massachusetts and the Infrared Processing and Analysis Center/California Institute of Technology, funded by the National Aeronautics and Space Administration and the National Science Foundation.

%Spitzer
This work is based [in part] on observations made with the Spitzer Space Telescope,
which is operated by the Jet Propulsion Laboratory, California Institute of Technology under
a contract with NASA. Support for this work was provided by NASA through an award issued
by JPL/Caltech.

%WISE
This publication makes use of data products from the Wide-field Infrared Survey
Explorer, which is a joint project of the University of California, Los Angeles,
and the Jet Propulsion Laboratory/California Institute of Technology,
funded by the National Aeronautics and Space Administration.

%VOSA
This publication makes use of VOSA, developed under the Spanish Virtual Observatory project supported from the Spanish MICINN through grant AyA2011-24052.

%TOPCAT
This work makes extensive use of TOPCAT and STILTS available
at http://www.starlink.ac.uk/topcat/ and http://www.starlink.ac.uk/stilts/

\facilities{Sloan (BOSS, Optical), ESO:VISTA, CTIO:2MASS, FLWO:2MASS, Spitzer (IRAC), WISE}

%\software{IRAF, IDL, VOSA, SPTCLASS, STILTS}

\appendix
\section{Field Stars}
\label{appendix}

The 119 objects resulting as field stars lack H$_\alpha$ emission and/or LiI$\lambda$6708 absorption, and show strong NaI$\lambda\lambda$8183, 8195 doublet in absorption. In Table \ref{tab:field_stars} we list these stars rejected as confirmed members as well as their spectral types together with their $I$ magnitudes and $I-J$ colors.

\begin{deluxetable}{ccccc}
\tabletypesize{\tiny}
\tablecaption{Stars on the BOSS plate rejected as confirmed members of 25 Ori or Orion OB1a.\label{tab:field_stars}}
\tablewidth{0pt}
\tablehead{
RA & DEC & SpT & I & I-J \\
   &     &     &   &
\vspace*{-1mm}
}
\decimalcolnumbers
\startdata
	80.842323 & 1.310326 & A4.8  $\pm$3.6   & 16.587 & .494  \\ \arrayrulecolor{Gray}\hline
	82.423676 & 1.507883 & A5.1  $\pm$3.9   & 15.071 & .562  \\ \arrayrulecolor{Gray}\hline
	81.696971 & 0.706731 & A5.2  $\pm$3.7   & 15.134 & .530  \\ \arrayrulecolor{Gray}\hline
	82.240343 & 1.305623 & A5.3  $\pm$4.5   & 15.242 & .568  \\ \arrayrulecolor{Gray}\hline
	80.048167 & 1.076809 & A6.4  $\pm$4.8   & 14.051 & .551  \\ \arrayrulecolor{Gray}\hline
	82.15245  & 0.754426 & A6.7  $\pm$4.2   & 14.655 & .565  \\ \arrayrulecolor{Gray}\hline
	82.527304 & 1.605351 & A8.2  $\pm$5.4   & 15.897 & .605  \\ \arrayrulecolor{Gray}\hline
	80.51536  & 0.695541 & F0$^a$$\pm$---   & 16.268 & .543  \\ \arrayrulecolor{Gray}\hline
	80.049821 & 1.307561 & F0$^a$$\pm$---   & 18.248 & .598  \\ \arrayrulecolor{Gray}\hline
	81.607394 & 1.125553 & F0.5  $\pm$4.5   & 16.802 & .499  \\ \arrayrulecolor{Gray}\hline
	79.928433 & 1.42041  & F1.8  $\pm$5.2   & 14.876 & ---   \\ \arrayrulecolor{Gray}\hline
	82.211707 & 1.089703 & F2.2  $\pm$4.9   & 15.305 & .508  \\ \arrayrulecolor{Gray}\hline
	82.112572 & 1.021283 & F2.3  $\pm$5.1   & 15.757 & .524  \\ \arrayrulecolor{black}
\enddata
\tabletypesize{\scriptsize}
\tablenotetext{a}{Spectral type assigned by the SDSS classification. Our SPTCLASS classification failed for this star.}
\tablecomments{The SDSS spectral type classification has not assigned the spectral type uncertainties. The complete version of this table is available in the electronic version of this publication.}
\end{deluxetable}

% \allauthors

\listofchanges

\end{document}